# Source Coding with Mismatched Distortion Measures

Urs Niesen, Devavrat Shah, and Gregory Wornell

**Abstract**

We consider the problem of lossy source coding with a mismatched distortion measure. That is, we investigate what distortion guarantees can be made with respect to distortion measure $\tilde{\rho}$, for a source code designed such that it achieves distortion less than $D$ with respect to distortion measure $\rho$. We find a single-letter characterization of this mismatch distortion and study properties of this quantity. These results give insight into the robustness of lossy source coding with respect to modeling errors in the distortion measure. They also provide guidelines on how to choose a good tractable approximation of an intractable distortion measure.

## I. Introduction

### A. Problem Formulation

Given source alphabet $\mathcal{X}$ and a reconstruction alphabet $\mathcal{Y}$, a source $\{X_i\}_{i\geq 1}$ with each $X_i$ taking values in $\mathcal{X}$, and two distortion measures $\rho_n, \tilde{\rho}_n : \mathcal{X}^n \times \mathcal{Y}^n \to \mathbb{R}_+$. Assume we have access to an oracle that, when queried, produces a source code $f_n$ (i.e., a mapping $f_n : \mathcal{X}^n \to \mathcal{Y}^n$) such that

$$\mathbb{E}\rho_n(X^n, f_n(X^n)) \leq D.$$

What guarantees can we make a priori (i.e., before querying the oracle) about $\mathbb{E}\tilde{\rho}_n(X^n, f_n(X^n))$? As a second question, assume we have access to an oracle that, when queried, produces a source code $f_n$ such that[1]

$$\frac{1}{n}\log|f_n(\mathcal{X}^n)| \leq R$$
$$\mathbb{E}\rho_n(X^n, f_n(X^n)) \leq D.$$

What guarantees can we make a priori about $\mathbb{E}\tilde{\rho}_n(X^n, f_n(X^n))$?

This problem has the following operational significance. Let a source code with expected distortion according to $\rho$ of at most $D$ be given. Assume instead of using this source code with respect to $\rho$, we decide to use it with respect to $\tilde{\rho}$. Such a situation occurs if constructing a source code for $\tilde{\rho}$ is not feasible or if $\tilde{\rho}$ is not fully known when constructing the source code. We are then faced with a mismatch in the distortion measure, and the best distortion guarantee mentioned in the opening paragraph provides a measure for how severe this mismatch is.

As an example, for an image compression problem, $\tilde{\rho}$ is determined by the human visual system, and any tractable model $\rho$ of it can necessarily be only an approximation of it. To be more specific, assume $\rho$ is taken to be squared error. While it is well known that this is not a faithful model for the human visual system, it is nevertheless often used in practice due to its simplicity. Assume then we choose one out of the many available source coding schemes for squared error distortion $\rho$. This source coding scheme will have some distortion guarantee for $\rho$ (the distortion measure it is designed for). The best performance guarantee mentioned in the opening paragraph allows then to translate this distortion guarantee for $\rho$ to a distortion guarantee for $\tilde{\rho}$. If, in addition, we also fix the rate of the source coding scheme, we are able to obtain a tighter performance guarantee (the second question in the opening paragraph).

This work was supported in part by NSF under Grant No. CCF-0515109, and by HP through the MIT/HP Alliance.

The authors are with the Massachusetts Institute of Technology, Department of Electrical Engineering and Computer Science, Cambridge, MA 02139, USA. Email: {uniesen,devavrat,gww}@mit.edu

[1] $|f_n(\mathcal{X}^n)|$ denotes the cardinality of the range of the function $f_n$.

In other words, an answer to the above questions allows to analyze the robustness of coding schemes to modeling errors - or mismatch in general - in the distortion measure.

### B. Related Work

The question of mismatched distortion measures in source coding has previously been considered in [1], [2], [3], [4], and [5]. In these works the mismatch is only with respect to the encoding part of the source code, whereas at least the decoder is matched to the proper distortion measure. This differs from the setup here, where the mismatch is with respect to both, the encoder and the decoder. We comment on the precise differences in the following paragraphs.

In [1], a partial order among distortion measures is defined such that $\rho \geq \tilde{\rho}$ if for every source code (consisting of an encoder $g_n : \mathcal{X}^n \to \{1, \dots, \exp(nR)\}$ and a decoder $\phi_n : \{1, \dots, \exp(nR)\} \to \mathcal{Y}^n$) satisfying $\mathbb{E}\rho_n(X^n, \phi_n(g_n(X^n))) \leq D$ there exists a second decoder $\tilde{\phi}_n$ satisfying $\mathbb{E}\tilde{\rho}_n(X^n, \tilde{\phi}_n(g_n(X^n))) \leq D$. Thus, in this setup, the encoder $g_n$ is designed for a mismatched distortion measure $\rho$, whereas the decoder $\tilde{\phi}_n$ is matched to the distortion measure $\tilde{\rho}$.

In [2], the following problem is considered. Fix a codebook $\mathcal{C} \subset \mathcal{Y}^n$, and let $g_n : \mathcal{X}^n \to \mathcal{C}$ be an optimal encoder for this codebook $\mathcal{C}$ with respect to $\rho$. Find codebook $\mathcal{C}$ and decoder $\tilde{\phi}_n : \mathcal{C} \to \mathcal{Y}^n$ such that $\mathbb{E}\tilde{\rho}_n(X^n, \tilde{\phi}_n(g_n(X^n)))$ is minimized. Again, the mismatch is only with respect to the encoder $g_n$, whereas the decoder as well as the codebook $\mathcal{C}$ are matched to the distortion measure $\tilde{\rho}$.

In [3], the author considers the problem of finding an encoder $g_n : \mathcal{X}^n \to \{1, \dots, \exp(nR)\}$ such that there exists a decoder $\phi_n : \{1, \dots, \exp(nR)\} \to \mathcal{Y}^n$ satisfying $\mathbb{E}\rho_n(X^n, \phi_n(g_n(X^n))) \leq D$ while maximizing $\inf_{\tilde{\phi}_n} \mathbb{E}\tilde{\rho}_n(X^n, \tilde{\phi}_n(g_n(X^n)))$. In other words, the goal is to find an encoder that guarantees distortion at most $D$ with respect to $\rho$, while making sure that this code has maximum possible distortion with respect to $\tilde{\rho}$. As in the previous cases, the mismatch is only with respect to the encoder, whereas the decoder $\tilde{\phi}_n$ is matched to the distortion measure $\tilde{\rho}$.

In [4, Problem 2.2.14] and [5], the problem of lossy source coding with respect to a class of distortion measures is considered: Given a class of distortion measures $\Gamma$, we want to find a source code $f_n : \mathcal{X}^n \to \mathcal{Y}^n$ such that $\sup_{\rho \in \Gamma} \mathbb{E}\rho_n(X^n, f_n(X^n))$ is minimized. In other words, $f_n$ is now "matched" to all $\rho \in \Gamma$ simultaneously.

### C. Modeling Perceptual Distortion Measures

In this section, we briefly review the typical structure of perceptual distortion measures. This will motivate the results presented in the main text. We focus here on distortion measures for image compression; the structure of perceptual distortion measures for speech, audio, or video compression is similar (see [6] for details on those distortion measures). The discussion here follows [7] and [8].

The typical structure of a perceptual distortion measure for image compression is depicted in Figure 1. Here $x$ and $y$ are the original and reconstructed image respectively, represented, for example, as vector of gray scale values.

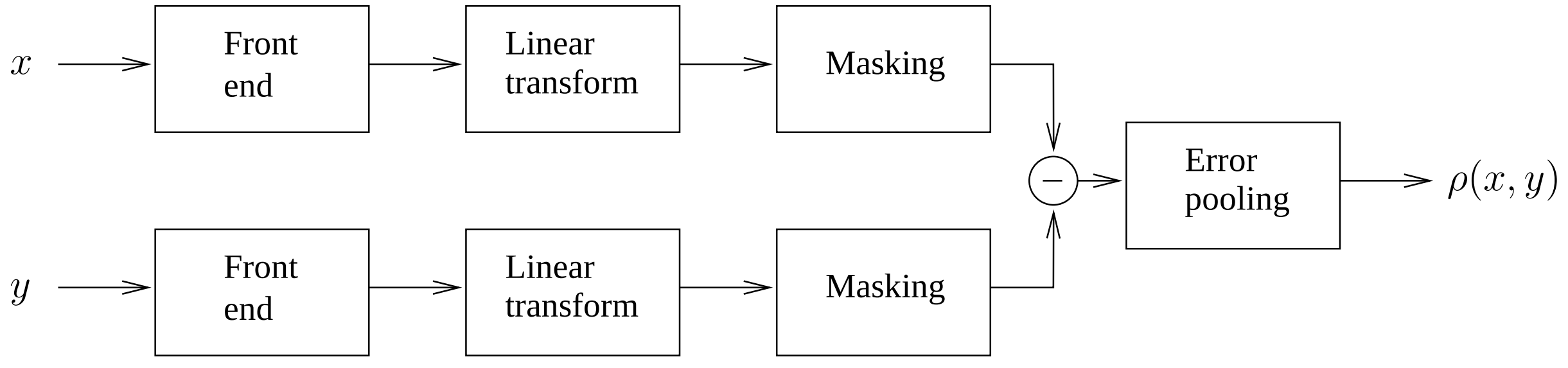

Fig. 1. Typical structure of a perceptual distortion measure. Adapted from [7].

The first block (termed front end) contains conversions from the image format to physical luminance observed by the human eye and other calibrations. The second block performs a linear transform of the two images, usually decomposing it into a number of spatial frequency bands with different orientations. In the next block, the coefficient of each band is weighted to account for masking effects. The resulting vector of weighted coefficients of the original and reconstructed image are then subtracted. The last block takes this vector of weighted differences and pools it together into one real number. Usually this is done by computing the $\ell_p$ norm of the difference vector for some $p \geq 1$ or taking some power $r \geq 1$ of that norm. Typical values of $p$ range from $2$ to $4$.

Formally, the source and reconstruction alphabets are $\mathcal{X} = \mathcal{Y} = \mathbb{R}^m$ or $\mathcal{X} = \mathcal{Y} = [0,1]^m$ for some finite $m$. In the following, we write $x, y$ for elements of general $\mathcal{X}$, $\mathcal{Y}$, and we write $\boldsymbol{x}, \boldsymbol{y}$ if we want to emphasize that $\mathcal{X} = \mathcal{Y} = \mathbb{R}^m$ or $\mathcal{X} = \mathcal{Y} = [0,1]^m$. This means that $\rho$ is of the form

$$\rho(\boldsymbol{x}, \boldsymbol{y}) = \big\| [v(x_1), \dots, v(x_m)] \boldsymbol{W}_{\boldsymbol{x}} - [v(y_1), \dots, v(y_m)] \boldsymbol{W}_{\boldsymbol{y}} \big\|_p^r,$$

and is sometimes simplified to

$$\rho(\boldsymbol{x}, \boldsymbol{y}) = \big\| \big( [v(x_1), \dots, v(x_m)] - [v(y_1), \dots, v(y_m)] \big) \boldsymbol{W}_{\boldsymbol{x}} \big\|_p^r. \tag{1}$$

$v : \mathbb{R} \to \mathbb{R}$ accounts for the front end, $\boldsymbol{W} : \mathbb{R}^m \to \mathbb{R}^{m \times k}$ accounts for the linear transform and masking. Here (and in the following), we write for $\boldsymbol{a} \in \mathbb{R}^k$ and $p \geq 1$

$$\|\boldsymbol{a}\|_p \triangleq \begin{cases} \big( \sum_{i=1}^k |a_i|^p \big)^{1/p} & \text{if } p < \infty, \\ \max_{1 \leq i \leq k} |a_i| & \text{if } p = \infty. \end{cases}$$

### *D. Outline of Results*

We now discuss several questions that arise when trying to construct and use perceptual distortion measures for source coding. These questions motivate the results presented in this paper, and they are used as examples throughout.

- The choice of $r$ and $p$ for the error pooling seems to vary quite considerably across different perceptual distortion measures for image compression. [9] uses $p = 2, r = 1$, [10] uses $p = 2.4, r = 1$, [11] uses $p = 4, r = 1$, and [12], [13] use $p = 2, r = 2$. It is therefore of interest to know how distortion mismatch in these two parameters affect the performance of the source code. This is discussed in Example 2 (using Theorems 1, 2, 3, 4).
- Given a class of distortion measures $\Gamma$, [12] suggests the following approach to find the "best" approximation $\rho \in \Gamma$ to the distortion measure implemented by the human visual system: Simulate the (information theoretically) optimal encoding scheme for all $\rho \in \Gamma$, and determine experimentally (i.e., by showing the original and distorted image to a human) the one yielding the smallest distortion. This optimal distortion measure is then declared to be the best approximation. While this approach yields indeed the best approximation $\rho \in \Gamma$ when used with the *optimal* infinite length source code, it is not clear a priori if this $\rho$ will also yield a good approximation when used with a *suboptimal* source code. Indeed, as we shall see in Example 2, there are situations in which the mismatch for the optimal and (even only slightly) suboptimal source codes are very different. In Example 3 (using Theorem 5), we provide conditions on $\Gamma$ and the source under which the $\rho$ found with this approach yields also a good approximation when used with good but not optimal source codes. These conditions hold for the model in [12] (with a few additional assumptions, that are implicitly made there). Hence our results provide evidence that the optimal approximation $\rho \in \Gamma$ found in [12] will also be good for practical (and hence necessarily suboptimal) source codes.
- [13] proposes a vector quantizer design procedure for distortion measures of the form

$$\rho(\boldsymbol{x}, \boldsymbol{y}) = w_{\boldsymbol{x}} \|\boldsymbol{y} - \boldsymbol{x}\|_2^2, \tag{2}$$

where $w : \mathbb{R}^m \to \mathbb{R}$. Since this is considerably simpler than the standard model (1), the question arises of how to find the $w_{\boldsymbol{x}}$ such that the resulting $\rho$ in (2) is "close" to one of the more complicated form (1). Note that it is not immediately obvious what "close" should mean in this context. Indeed, there are several such notions that are reasonable. In Example 5, we show what properties such a notion should have. The problem posed by [13] discussed above is treated in detail in Example 1 (using Theorem 3) and Example 6 (using Corollaries 8 and 9).

- Essentially all models of perceptual distortion measures contain a number of parameters that are usually chosen to be in "close agreement" with the behavior of the human visual system. Again, it is not clear what "close agreement" should mean here. In Example 7 (using Proposition 10), a simple such measure of closeness is proposed, providing a guideline for how to tune the parameters of a perceptual distortion model to be used for source coding.

### *E. Organization*

The remainder of this paper is organized as follows. In Section II, we present our main results. Section III contains the corresponding proofs. Section IV contains concluding remarks.

## II. Main Results

In this section, we formally introduce the problem of source coding with distortion mismatch. To simplify the exposition, and since it represents the case of most practical interest, we assume in the following that $\mathcal{X} = \mathcal{Y} = \mathbb{R}^m$ for some finite $m$. Most of the results are, however, also valid if the alphabets are general Polish spaces (i.e., complete, separable, metric spaces). We let $\mathcal{B}(\mathcal{X} \times \mathcal{Y})$ be the Borel sets of $\mathcal{X} \times \mathcal{Y}$. By $\mathcal{P}(\mathcal{X} \times \mathcal{Y})$, we denote the set of all probability measures on $(\mathcal{X} \times \mathcal{Y}, \mathcal{B}(\mathcal{X} \times \mathcal{Y}))$. For $Q \in \mathcal{P}(\mathcal{X} \times \mathcal{Y})$, $Q_{\mathcal{X}}$ denotes the $\mathcal{X}$ marginal of $Q$. For a measurable function $g : \mathcal{X} \times \mathcal{Y} \to \mathbb{R}$, we denote by $\mathbb{E}_Q g(X, Y)$ or $\mathbb{E}_Q g$ the expectation of $g(X, Y)$ with respect to $Q$. For any $A \in \mathcal{B}(\mathcal{X} \times \mathcal{Y})$, we write $\mathbb{E}_Q(g; A)$ for $\mathbb{E}_Q g \mathbb{1}_A$. $I(Q)$ denotes mutual information (in nats) between the random variables $(X, Y) \sim Q$. Throughout this paper, we restrict attention to single-letter distortion measures, i.e., measurable functions $\rho : \mathcal{X} \times \mathcal{Y} \to \mathbb{R}_+$ with $\rho_n : \mathcal{X}^n \times \mathcal{Y}^n \to \mathbb{R}_+$ defined by

$$\rho_n(x^n, y^n) = \frac{1}{n} \sum_{i=1}^{n} \rho(x_i, y_i).$$

We also assume throughout that the source $\{X_i\}_{i \geq 1}$ is i.i.d. with distribution $P \in \mathcal{P}(\mathcal{X})$. $\mathsf{R}_\rho(D)$ and $\mathsf{D}_\rho(R)$ denote the rate-distortion and the distortion-rate function for the source $\{X_i\}_{i \geq 1}$ and with respect to the single-letter distortion measure $\rho$, i.e.,

$$\mathsf{R}_\rho(D) \triangleq \inf_{\substack{Q \in \mathcal{P}(\mathcal{X} \times \mathcal{Y}): \\ Q_{\mathcal{X}} = P, \mathbb{E}_Q \rho \leq D}} I(Q),$$

$$\mathsf{D}_\rho(R) \triangleq \inf_{\substack{Q \in \mathcal{P}(\mathcal{X} \times \mathcal{Y}): \\ Q_{\mathcal{X}} = P, I(Q) \leq R}} \mathbb{E}_Q \rho.$$

Our results are divided into several parts. In Section II-A, we provide single-letter characterizations of the mismatch distortion. In Section II-B, we investigate properties of these quantities. Section II-C contains information on how to evaluate the single-letter characterizations of the mismatch distortion. Section II-D, considers the problem of finding a good representation of a distortion measure from a class of simpler ones.

### A. Single-Letter Characterizations

In this section, we provide single-letter characterizations of the smallest distortion with respect to $\tilde{\rho}$ that can be guaranteed for any source code (either with or without constraint on the rate $R$) designed for distortion $D_\rho$ with respect to $\rho$.

Define

$$\mathsf{D}_{\rho,\tilde{\rho}}(R, D_\rho) \triangleq \sup \mathbb{E}_Q \tilde{\rho}, \tag{3}$$

where the supremum is taken over all $Q \in \mathcal{P}(\mathcal{X} \times \mathcal{Y})$ such that $Q_{\mathcal{X}} = P$, $\mathbb{E}_Q \rho \le D_\rho$ and $I(Q) \le R$. If the set over which this supremum is taken is empty, we define $\mathsf{D}_{\rho,\tilde{\rho}}(R, D_\rho) \triangleq -\infty$.

**Theorem 1.** *Let $\rho, \tilde{\rho}$ be distortion measures satisfying $\mathbb{E}_P \rho(X, y_0) < \infty$ for some $y_0 \in \mathcal{Y}$. For every $D_{\tilde{\rho}} < \infty$ such that*

$$0 \le D_{\tilde{\rho}} < \lim_{\delta \downarrow 0} \mathsf{D}_{\rho,\tilde{\rho}}(R - \delta, D_\rho - \delta)$$

*there exists a sequence of source codes $\{f_n\}_{n \ge 1}$ such that*

$$\lim_{n\to\infty} \frac{1}{n} \log |f_n(\mathcal{X}^n)| \le R,$$
$$\limsup_{n\to\infty} \mathbb{E}\rho_n(X^n, f_n(X^n)) \le D_\rho,$$
$$\liminf_{n\to\infty} \mathbb{E}\tilde{\rho}_n(X^n, f_n(X^n)) \ge D_{\tilde{\rho}}.$$

**Theorem 2.** *For any $n$ and any source code $f_n : \mathcal{X}^n \to \mathcal{Y}^n$ such that*

$$\frac{1}{n} \log |f_n(\mathcal{X}^n)| = R,$$
$$\mathbb{E}\rho_n(X^n, f_n(X^n)) \le D_\rho,$$

*we have*[2]

$$\mathbb{E}\tilde{\rho}_n(X^n, f_n(X^n)) \le \mathsf{D}_{\rho,\tilde{\rho}}(R+, D_\rho).$$

*If, moreover, $R > \mathsf{R}_\rho(D_\rho)$ then*

$$\mathbb{E}\tilde{\rho}_n(X^n, f_n(X^n)) \le \mathsf{D}_{\rho,\tilde{\rho}}(R, D_\rho).$$

Theorems 1 and 2 allow us to make guarantees about the performance of a source code constructed with mismatched distortion measure. Indeed, if $f_n : \mathcal{X}^n \to \mathcal{Y}^n$ is a source code of rate $R$ designed for a distortion measure $\rho$ and distortion level $D_\rho$, then by Theorem 2, $f_n$ is also a source code for any distortion measure $\tilde{\rho}$ and distortion level $\mathsf{D}_{\rho,\tilde{\rho}}(R+, D_\rho)$. Moreover, this is essentially the best guarantee one can make, since by Theorem 1 there exist source codes with same blocklength $n$ and same rate $R$ designed for distortion measure $\rho$ and distortion level $D_\rho$ that result in a distortion level of more than

$$\mathsf{D}_{\rho,\tilde{\rho}}(R - \delta(n), D_\rho - \delta(n)) - \delta(n)$$

for distortion measure $\tilde{\rho}$ with $\delta(n) \to 0$ as $n \to \infty$. This answers the second question posed in the introduction.

To answer the first question, we need to find the best distortion guarantee that is independent of the rate $R$ of the source code. From Theorems 1 and 2, this best distortion guarantee is given by

$$\sup_{R \ge 0} \mathsf{D}_{\rho,\tilde{\rho}}(R, D_\rho).$$

[2]For a real valued function $g$, we write $g(x+) \triangleq \lim_{\delta\downarrow 0} g(x + \delta)$ and $g(x-) \triangleq \lim_{\delta\downarrow 0} g(x - \delta)$, assuming the limits exist.

Since $\mathsf{D}_{\rho,\tilde{\rho}}(\cdot, D_\rho)$ is an increasing function, this is equal to

$$\lim_{R\to\infty} \mathsf{D}_{\rho,\tilde{\rho}}(R, D_\rho).$$

The next theorem considers this limit.

**Theorem 3.** *If*

(i) $\rho, \tilde{\rho}$ *are continuous*

(ii) *there exists* $y_0 \in \mathcal{Y}$ *such that* $\mathbb{E}_P \rho(X, y_0) < \infty$

(iii) $\mathsf{D}_\rho(\infty) < D_\rho < \infty$

*then for any* $\eta \geq 0$ *the expectation*

$$\mathbb{E}_P \sup_{y\in\mathcal{Y}} (\tilde{\rho}(X, y) - \eta\rho(X, y))$$

*is well defined and*

$$\mathsf{D}_{\rho,\tilde{\rho}}(\infty, D_\rho) = \min_{\eta\geq 0} \Big(\eta D_\rho + \mathbb{E}_P \sup_{y\in\mathcal{Y}} \big(\tilde{\rho}(X, y) - \eta\rho(X, y)\big)\Big).$$

*If, moreover,*

(iv) $\mathsf{D}_{\rho,\tilde{\rho}}(\infty, D_\rho) < \infty$,

*then*

$$\lim_{R\to\infty} \mathsf{D}_{\rho,\tilde{\rho}}(R, D_\rho) = \mathsf{D}_{\rho,\tilde{\rho}}(\infty, D_\rho).$$

**Example 1.** Let

$$\rho(\boldsymbol{x}, \boldsymbol{y}) = (\boldsymbol{y} - \boldsymbol{x})^T \boldsymbol{W_x} (\boldsymbol{y} - \boldsymbol{x}),$$
$$\tilde{\rho}(\boldsymbol{x}, \boldsymbol{y}) = (\boldsymbol{y} - \boldsymbol{x})^T \widetilde{\boldsymbol{W}}_{\boldsymbol{x}} (\boldsymbol{y} - \boldsymbol{x}),$$

where $\boldsymbol{W_x}$ and $\widetilde{\boldsymbol{W}}_{\boldsymbol{x}}$ are positive definite for $P$ almost every $\boldsymbol{x}$. Let $P \in \mathcal{P}(\mathcal{X})$ such that

$$\mathbb{E}_P \boldsymbol{X}^T \boldsymbol{W_X} \boldsymbol{X} < \infty.$$

With this, Assumption (i) and (ii) of Theorem 3 are satisfied. Applying the theorem yields that for $\mathsf{D}_\rho(\infty) < D_\rho < \infty$,

$$\mathsf{D}_{\rho,\tilde{\rho}}(\infty, D_\rho) = \min_{\eta\geq 0} \eta D_\rho + \mathbb{E}_P \sup_{\boldsymbol{y}\in\mathbb{R}^m} (\boldsymbol{y} - \boldsymbol{X})^T (\widetilde{\boldsymbol{W}}_{\boldsymbol{X}} - \eta \boldsymbol{W_X})(\boldsymbol{y} - \boldsymbol{X}), \tag{4}$$

and whenever this quantity is finite then also

$$\lim_{R\to\infty} \mathsf{D}_{\rho,\tilde{\rho}}(R, D_\rho) = \mathsf{D}_{\rho,\tilde{\rho}}(\infty, D_\rho).$$

If $\widetilde{\boldsymbol{W}}_{\boldsymbol{x}} - \eta \boldsymbol{W_x}$ in (4) is not negative semidefinite for some $\boldsymbol{x}$, then it has at least one strictly positive eigenvalue $\nu > 0$ with corresponding eigenvector $\boldsymbol{v}$. Setting $\boldsymbol{y} = \boldsymbol{x} - a\boldsymbol{v}$ yields

$$(\boldsymbol{y} - \boldsymbol{x})^T (\widetilde{\boldsymbol{W}}_{\boldsymbol{x}} - \eta \boldsymbol{W_x})(\boldsymbol{y} - \boldsymbol{x}) = a^2 \nu \boldsymbol{v}^T \boldsymbol{v} \to \infty$$

as $a \to \infty$. Hence the $\eta$ minimizing (4) is always such that $\widetilde{\boldsymbol{W}}_{\boldsymbol{x}} - \eta \boldsymbol{W_x}$ is negative semidefinite for $P$ almost every $\boldsymbol{x}$. In this case

$$\sup_{\boldsymbol{y}\in\mathbb{R}^m} (\boldsymbol{y} - \boldsymbol{x})^T (\widetilde{\boldsymbol{W}}_{\boldsymbol{x}} - \eta \boldsymbol{W_x})(\boldsymbol{y} - \boldsymbol{x}) = 0,$$

and we obtain

$$\lim_{R\to\infty} \mathsf{D}_{\rho,\tilde{\rho}}(R, D_\rho) = D_\rho \inf\{\eta \geq 0 : \widetilde{\boldsymbol{W}}_{\boldsymbol{x}} - \eta \boldsymbol{W_x} \leq 0 \ P \text{ a.e.}\}, \tag{5}$$

where $\widetilde{\boldsymbol{W}}_{\boldsymbol{x}} - \eta \boldsymbol{W_x} \leq 0$ means that the matrix on the left hand side is negative semidefinite. $\Diamond$

### B. Properties of $\mathsf{D}_{\rho,\tilde{\rho}}(R, D_\rho)$

The function $\mathsf{D}_{\rho,\tilde{\rho}}(R, D_\rho)$ exhibits the following behavior:

$$\mathsf{D}_{\rho,\tilde{\rho}}(R, D_\rho) \in \begin{cases} \{-\infty\} & \text{if } R < \mathsf{R}_\rho(D_\rho), \\ \mathbb{R}_+ \cup \{\pm\infty\} & \text{if } R = \mathsf{R}_\rho(D_\rho), \\ \mathbb{R}_+ \cup \{\infty\} & \text{if } R > \mathsf{R}_\rho(D_\rho). \end{cases}$$

Moreover, a simple argument shows that $\mathsf{D}_{\rho,\tilde{\rho}}(R, D_\rho)$ is concave and increasing in both its arguments, and continuous at all points $(R, D_\rho)$ such that $R > \mathsf{R}_\rho(D_\rho)$. $\mathsf{D}_{\rho,\tilde{\rho}}(R, D_\rho)$ is necessarily discontinuous at $(\mathsf{R}_\rho(D_\rho), D_\rho)$, but could be either left- or right-continuous (as a function of either $R$ or $D_\rho$). This implies that the function either equals $\infty$ for all $(R, D_\rho)$ such that $R > \mathsf{R}_\rho(D_\rho)$ or is finite on this whole range. The two types of possible behaviors of $\mathsf{D}_{\rho,\tilde{\rho}}(R, D_\rho)$ are depicted in Figure 2.

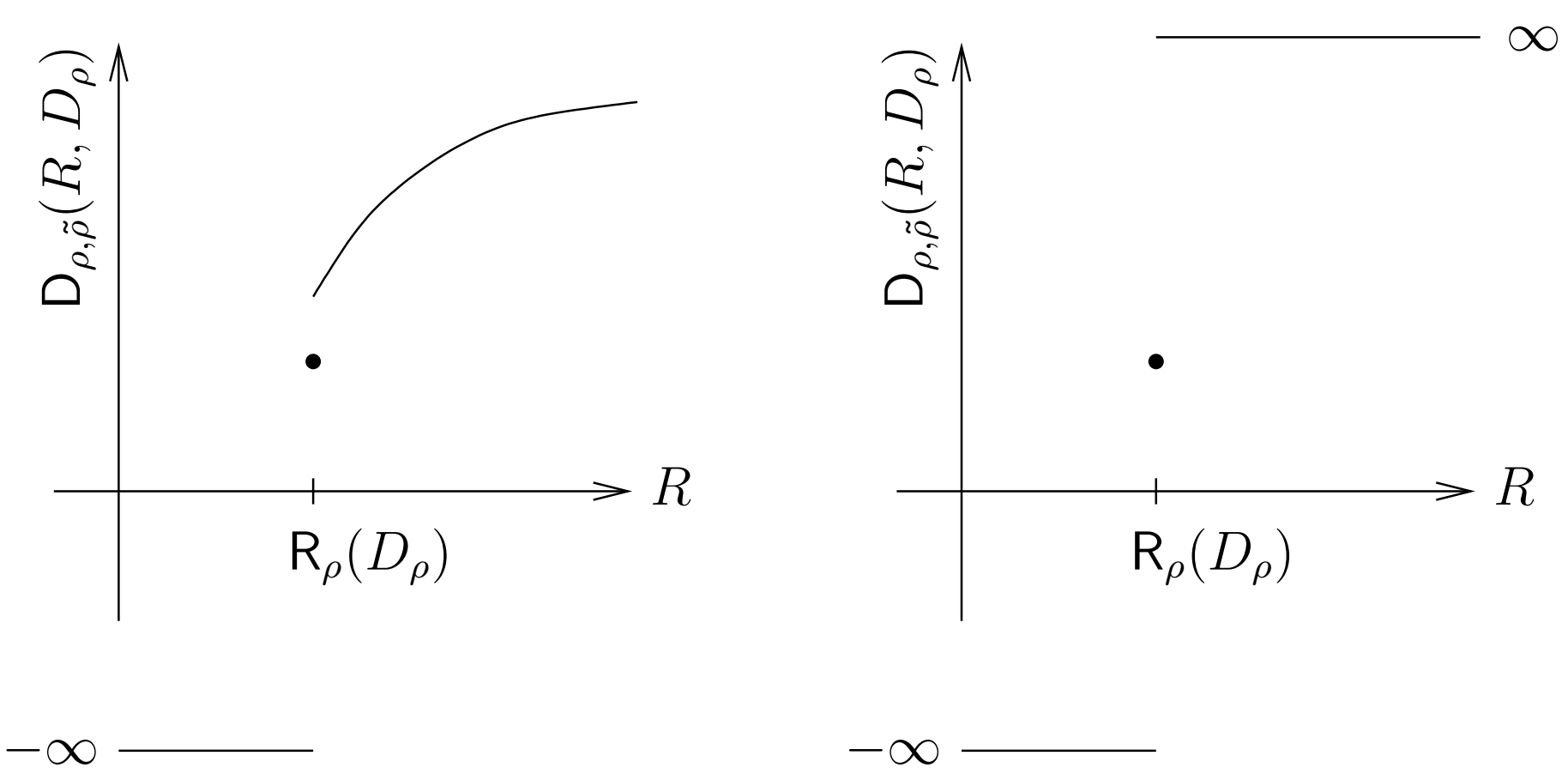

Fig. 2. Possible behaviors of $\mathsf{D}_{\rho,\tilde{\rho}}(R, D_\rho)$.

The next two theorems describe the behavior of $\mathsf{D}_{\rho,\tilde{\rho}}(R, D_\rho)$ in more detail. Theorem 4 provides conditions under which $\mathsf{D}_{\rho,\tilde{\rho}}(R, D_\rho) = \infty$ for all $(R, D_\rho)$ such that $R > \mathsf{R}_\rho(D_\rho)$. In these situations, we cannot make any guarantees about the performance of a source code of rate $R$ designed for distortion measure $\rho$ and distortion level $D_\rho$ when used for distortion measure $\tilde{\rho}$. Theorem 5 gives sufficient conditions such that $\mathsf{D}_{\rho,\tilde{\rho}}(R, D_\rho) \geq 0$ for $(R, D_\rho)$ with $R = \mathsf{R}_\rho(D_\rho)$, and conditions for $\mathsf{D}_{\rho,\tilde{\rho}}(R, D_\rho)$ to be right-continuous in $R$ at those points.

**Theorem 4.** *If*

(i) $0 < R < \infty$
(ii) $D_\rho > \mathsf{D}_\rho(R)$
(iii) *there exists* $y_0 \in \mathcal{Y}$ *such that* $\mathbb{E}_P \rho(X, y_0) \triangleq D_0 < \infty$
(iv) *there exist* $\{A_k\}_{k\geq 1} \subset \mathcal{B}(\mathcal{X})$, $\{y_k^*\}_{k\geq 1} \subset \mathcal{Y}$ *such that*

$$\begin{aligned} &\mathbb{E}_P(\rho(X, y_k^*); A_k) < \infty && \text{for all } k \geq 1, \\ &P(A_k) \inf_{x\in A_k} \tilde{\rho}(x, y_k^*) \to \infty && \text{as } k \to \infty, \\ &\sup_{x\in A_k} \rho(x, y_k^*)/\tilde{\rho}(x, y_k^*) \to 0 && \text{as } k \to \infty \end{aligned}$$

*then* $\mathsf{D}_{\rho,\tilde{\rho}}(R, D_\rho) = \infty$.

*Remark.* The second and third part of Assumption (iv) are satisfied for example if $\tilde{\rho}(x, y) \to \infty$ and $\rho(x, y)/\tilde{\rho}(x, y) \to 0$ when $\|y - x\|_2 \to \infty$. See also Example 2.

**Example 2.** Let $\rho(x,y) = d(y-x)^r$, $\tilde{\rho}(x,y) = \tilde{d}(y-x)^{\tilde{r}}$ for arbitrary norms $d, \tilde{d} : \mathbb{R}^m \to \mathbb{R}_+$, and for $r, \tilde{r} \geq 1$. Let $P \in \mathcal{P}(X)$ be such that $\mathbb{E}_P d(X)^r < \infty$. With slight abuse of notation, we shall write $\rho(x-y)$ for $\rho(x,y)$ and similar for $\tilde{\rho}$ in this example.

*Case 1:* $r < \tilde{r}$. We first show that the conditions of Theorem 4 are satisfied. Since all norms on a finite dimensional space are equivalent, there exist $a_1, a_2 > 0$ such that

$$a_1 d(z) \leq \tilde{d}(z) \leq a_2 d(z)$$

for all $z \in \mathbb{R}^m$, and thus there exist $b_1, b_2 > 0$ such that

$$b_1 \rho(x-y)^{\tilde{r}/r} \leq \tilde{\rho}(x-y) \leq b_2 \rho(x-y)^{\tilde{r}/r}$$

for all $x \in \mathcal{X}, y \in \mathcal{Y}$. Hence, we have

$$\rho(x-y)/\tilde{\rho}(x-y) \leq \frac{1}{b_1} \rho(x-y)^{(r-\tilde{r})/r}$$

for all $x \in \mathcal{X}, y \in \mathcal{Y}$. Let $A \triangleq [-c,c]^m$, and choose $c$ such that $P(A) > 0$. Set $y_k^* \triangleq k\mathbf{1}$, where $\mathbf{1} = (1,\ldots,1) \in \mathbb{R}^m$. With this

$$\begin{aligned}\sup_{x\in A} \rho(x-y_k^*)/\tilde{\rho}(x-y_k^*) &\leq \sup_{x\in A} \frac{1}{b_1} \rho(x-y_k^*)^{(r-\tilde{r})/r} \\ &= \max_{x\in A} \frac{1}{b_1} d(x-y_k^*)^{r-\tilde{r}} \to 0\end{aligned}$$

as $k \to \infty$, satisfying Assumption (iv.3) of Theorem 4. Moreover,

$$P(A) \inf_{x\in A} \tilde{\rho}(x-y_k^*) = P(A) \min_{x\in A} \tilde{d}(x-y_k^*)^{\tilde{r}} \to \infty$$

as $k \to \infty$, satisfying Assumption (iv.2) of. Finally,

$$\mathbb{E}_P \rho(X-y_k^*) \leq \mathbb{E}_P \big(d(y_k^*) + d(X)\big)^r.$$

By Jensen's inequality

$$\Big(\frac{1}{2} d(y_k^*) + \frac{1}{2} d(X)\Big)^r \leq \frac{1}{2} d(y_k^*)^r + \frac{1}{2} d(X)^r,$$

and hence

$$\mathbb{E}_P \rho(X, y_k^*) \leq 2^{r-1} \big(d(y_k^*)^r + \mathbb{E}_P d(X)^r\big) \leq \infty$$

for all $k \geq 0$. Therefore with $y_0 = 0$, we have $\mathbb{E}_P \rho(X-y_0) < \infty$ and $\mathbb{E}_P(\rho(X-y_k^*); A) < \infty$, satisfying Assumptions (iii) and (iv.1) of Theorem 4. Thus applying the Theorem with $A_k \triangleq A$ yields

$$\mathsf{D}_{\rho,\tilde{\rho}}(R, D_\rho) = \infty$$

for all $0 < R < \infty$ and $D_\rho > \mathsf{D}_\rho(R)$.

*Case 2:* $r = \tilde{r}$. Clearly $\rho$ and $\tilde{\rho}$ are continuous, and $\mathbb{E}_P \rho(X) < \infty$. Hence Theorem 3 asserts that for $\mathsf{D}_\rho(\infty) < D_\rho < \infty$

$$\begin{aligned}\mathsf{D}_{\rho,\tilde{\rho}}(\infty, D_\rho) &= \min_{\eta\geq 0} \Big(\eta D_\rho + \mathbb{E}_P \sup_{y\in\mathcal{Y}} \big(\tilde{\rho}(X,y) - \eta\rho(X,y)\big)\Big) \\ &= \min_{\eta\geq 0} \Big(\eta D_\rho + \sup_{z\in\mathbb{R}^m} \tilde{\rho}(z) - \eta\rho(z)\Big),\end{aligned} \tag{6}$$

and that this quantity is equal to $\lim_{R\to\infty} \mathsf{D}_{\rho,\tilde{\rho}}(R, D_\rho)$ whenever it is finite.

Set

$$v^* \in \arg\max_{v\in\mathbb{R}^m : d(v)=1} \tilde{d}(v).$$

Since $\tilde{d}$ is continuous and $\{v : d(v) = 1\}$ is compact, at least one such maximizer exists. It is easy to check that

$$\sup_{z\in\mathbb{R}^m} \tilde{\rho}(z) - \eta\rho(z) = \sup_{a\geq 0} a^{\tilde{r}}\tilde{d}(v^*)^{\tilde{r}} - \eta a^r = \sup_{a\geq 0} a^r(\tilde{d}(v^*)^r - \eta), \tag{7}$$

where we have used $r = \tilde{r}$. In other words, the maximizing $z$ is of the form $av^*$ for some $a \geq 0$. If $\eta < \tilde{d}(v^*)^r$ then

$$\sup_{a\geq 0} a^r(\tilde{d}(v^*)^r - \eta) = \lim_{a\to\infty} a^r(\tilde{d}(v^*)^r - \eta) = \infty.$$

On the other hand, if $\eta \geq \tilde{d}(v^*)^r$, then

$$\sup_{a\geq 0} a^r(\tilde{d}(v^*)^r - \eta) = \lim_{a\to 0} a^r(\tilde{d}(v^*)^r - \eta) = 0.$$

Therefore the minimizing $\eta \geq 0$ in (6) is equal to $\tilde{d}(v^*)^r$ and

$$\lim_{R\to\infty} \mathsf{D}_{\rho,\tilde{\rho}}(R, D_\rho) = D_\rho \tilde{d}(v^*)^r.$$

*Case 3:* $r > \tilde{r}$. Recall that by (7)

$$\sup_{a\geq 0}\sup_{z\in\mathbb{R}^m} \tilde{\rho}(z) - \eta\rho(z) = a^{\tilde{r}}\tilde{d}(v^*)^{\tilde{r}} - \eta a^r$$

The optimal $a^* \geq 0$ maximizing this quantity is

$$a^* = \left(\frac{\tilde{r}}{\eta r}\tilde{d}(v^*)^{\tilde{r}}\right)^{1/(r-\tilde{r})},$$

which by Theorem 3 implies that for $\mathsf{D}_\rho(\infty) < D_\rho < \infty$

$$\mathsf{D}_{\rho,\tilde{\rho}}(\infty, D_\rho) = \min_{\eta\geq 0} \eta D_\rho + \eta^{-\tilde{r}/(r-\tilde{r})} b \triangleq \min_{\eta\geq 0} g(\eta),$$

where

$$b \triangleq \tilde{d}(v^*)^{\tilde{r}r/(r-\tilde{r})}\left(\left(\frac{\tilde{r}}{r}\right)^{\tilde{r}/(r-\tilde{r})} - \left(\frac{\tilde{r}}{r}\right)^{r/(r-\tilde{r})}\right) > 0.$$

The $\eta^*$ minimizing $g$ is

$$\eta^* = \left(\frac{r-\tilde{r}}{b\tilde{r}} D_\rho\right)^{(\tilde{r}-r)/r},$$

which finally yields

$$\lim_{R\to\infty} \mathsf{D}_{\rho,\tilde{\rho}}(R, D_\rho) = D_\rho^{\tilde{r}/r}\left(\frac{b}{r-\tilde{r}}\right)^{(r-\tilde{r})/r}\tilde{r}^{-\tilde{r}/r}r.$$

For $d = \tilde{d}$, $r = 2$, $\tilde{r} = 1$, this reduces to

$$\lim_{R\to\infty} \mathsf{D}_{\rho,\tilde{\rho}}(R, D_\rho) = \sqrt{D_\rho}.$$

◊

Theorem 4 characterizes the behavior of $\mathsf{D}_{\rho,\tilde{\rho}}(R, D_\rho)$ for $(R, D_\rho)$ such that $R > \mathsf{R}_\rho(D_\rho)$. The next theorem characterizes the behavior of $\mathsf{D}_{\rho,\tilde{\rho}}(R, D_\rho)$ for $(R, D_\rho)$ such that $R = \mathsf{R}_\rho(D_\rho)$.

**Theorem 5.** *Let the distortion measure $\rho$ be continuous, and $D_\rho > 0$. If there exist compact sets $K_k \subset \mathcal{X}$, $M_k \subset \mathcal{Y}$ such that $P(K_k) \to 1$ as $k \to \infty$ and*

$$\inf_{x\in K_k, y\in M_k^c} \rho(x,y) \to \infty \tag{8}$$

*as $k \to \infty$, then $\mathsf{D}_{\rho,\tilde{\rho}}(\mathsf{R}_\rho(D_\rho), D_\rho) \geq 0$, i.e., the set over which we optimize in* (3) *is non-empty.*

*If, in addition,* $\mathsf{D}_{\rho,\tilde{\rho}}(\mathsf{R}_\rho(D_\rho)+r, D_\rho) < \infty$ *for some* $r > 0$, $\tilde{\rho}$ *is continuous, and there exists* $a > 1$ *and* $c \geq 0$ *such that* $\tilde{\rho}^a \leq c + \rho$*, then*

$$\mathsf{D}_{\rho,\tilde{\rho}}(\mathsf{R}_\rho(D_\rho)+, D_\rho) = \mathsf{D}_{\rho,\tilde{\rho}}(\mathsf{R}_\rho(D_\rho), D_\rho).$$

*Remark.* Condition (8) is satisfied for example for $\rho$ such that $\rho(x, y) \to \infty$ as $\|y - x\|_2 \to \infty$. Indeed, for $K_k = [-k, k]^m$ and $M_k = [-2k, 2k]^m$,

$$\lim_{k \to \infty} P(K_k) = 1,$$

and

$$\inf_{x \in K_k, y \in M_k^c} \rho(x, y) \geq \inf_{x,y: \|y-x\|_2 \geq k} \rho(x, y) \to \infty$$

as $k \to \infty$.

**Example 3.** Given a class of distortion measures $\Gamma$, the following approach is suggested in [12] to find the "closest" one to $\tilde{\rho}$ implemented by the human visual system: Determine $\mathsf{D}_{\rho,\tilde{\rho}}(R, \mathsf{D}_\rho(R))$ for each $\rho \in \Gamma$ and pick a minimizer $\rho^*$. In situations where a unique distribution $Q$ with $Q_{\mathcal{X}} = P$ achieving $\mathsf{D}_\rho(R)$ exists, $\mathsf{D}_{\rho,\tilde{\rho}}(R, \mathsf{D}_\rho(R))$ can be found empirically by generating samples from $Q$ and having them evaluated by human subjects. The hope is that the distortion measure minimizing $\mathsf{D}_{\rho,\tilde{\rho}}(R, \mathsf{D}_\rho(R))$ should be a good approximation to $\tilde{\rho}$ also for non-optimal image compression schemes. Formally, this amounts to assuming that $\mathsf{D}_{\rho,\tilde{\rho}}(R+r, \mathsf{D}_\rho(R))$ is close to $\mathsf{D}_{\rho,\tilde{\rho}}(R, \mathsf{D}_\rho(R))$ (at least for small $r$). Hence this approach is only valid, if $\mathsf{D}_{\rho,\tilde{\rho}}(R+r, \mathsf{D}_\rho(R))$ is right continuous in $r$ at $r = 0$.

Theorem 5 gives conditions under which this is indeed the case. In [12], $\mathcal{X} = \mathcal{Y} = \mathbb{R}_+^m$, and each $\rho \in \Gamma$ is of the form

$$\rho(\boldsymbol{x}, \boldsymbol{y}) = \big\| \big( [v(x_1), \dots, v(x_m)] - [v(y_1), \dots, v(y_m)] \big) \boldsymbol{W} \big\|_2^2$$

for some monotonic increasing concave function $v : \mathbb{R}_+ \to \mathbb{R}$ and some matrix $\boldsymbol{W} \in \mathbb{R}^{m \times m}$. In order to apply Theorem 5, we need the additional assumptions that $v$ is continuous at $0$, that $v(s) \to \infty$ as $s \to \infty$, that $\boldsymbol{W}^T \boldsymbol{W}$ is positive definite, and that $\tilde{\rho}$ implemented by the human visual system is continuous and bounded. From Theorem 5, we obtain that under these conditions — implicitly made in [12] — $\mathsf{D}_{\rho,\tilde{\rho}}(R+r, \mathsf{D}_\rho(R))$ is indeed right continuous at $r = 0$, showing that $\rho^*$ should yield a good approximation to $\tilde{\rho}$ also for compression schemes that are only close to optimal.

We consider the problem of finding an optimal $\rho \in \Gamma$ approximating a given $\tilde{\rho}$ in more detail in Section II-D. $\Diamond$

### *C. Computing* $\mathsf{D}_{\rho,\tilde{\rho}}(R, D_\rho)$

Define

$$\mathsf{R}_{\rho,\tilde{\rho}}(D_\rho, D_{\tilde{\rho}}) \triangleq \inf I(Q),$$

where the infimum is taken over all $Q \in \mathcal{P}(\mathcal{X} \times \mathcal{Y})$ such that $Q_{\mathcal{X}} = P$, $\mathbb{E}_Q \rho \leq D_\rho$ and $\mathbb{E}_Q \tilde{\rho} \geq D_{\tilde{\rho}}$. Setting

$$\mathcal{S}_1 \triangleq \big\{ (R, D_\rho, D_{\tilde{\rho}}) : D_{\tilde{\rho}} \leq \mathsf{D}_{\rho,\tilde{\rho}}(R, D_\rho) \big\},$$
$$\mathcal{S}_2 \triangleq \big\{ (R, D_\rho, D_{\tilde{\rho}}) : R \geq \mathsf{R}_{\rho,\tilde{\rho}}(D_\rho, D_{\tilde{\rho}}) \big\},$$

it is easy to show that the closures of $\mathcal{S}_1$ and $\mathcal{S}_2$ are identical. It is convenient in the following to analyze $\mathsf{R}_{\rho,\tilde{\rho}}(D_\rho, D_{\tilde{\rho}})$ instead of $\mathsf{D}_{\rho,\tilde{\rho}}(R, D_\rho)$.

Define

$$\mathcal{Q}_1(D_\rho, D_{\tilde{\rho}}) \triangleq \{ Q \in \mathcal{P}(\mathcal{X} \times \mathcal{Y}) : Q_{\mathcal{X}} = P, \mathbb{E}_Q \rho \leq D_\rho, \mathbb{E}_Q \tilde{\rho} \geq D_{\tilde{\rho}} \},$$
$$\mathcal{Q}_2(D_\rho, D_{\tilde{\rho}}) \triangleq \{ Q \in \mathcal{Q}_1(D_\rho, D_{\tilde{\rho}}) : Q \ll \lambda_{\mathbb{R}^m \times \mathbb{R}^m} \},$$

where $\lambda_{\mathbb{R}^m\times\mathbb{R}^m}$ is Lebesgue measure on $\mathbb{R}^m \times \mathbb{R}^m$. Note that if $Q \ll \lambda_{\mathbb{R}^m\times\mathbb{R}^m}\}$, i.e., $Q$ is absolutely continuous with respect to Lebesgue measure, then $Q$ admits a density.

The next theorem gives conditions under which we can restrict the minimization in the definition of $\mathsf{R}_{\rho,\tilde{\rho}}(D_\rho, D_{\tilde{\rho}})$ to distributions admitting a density. We then use this result to find tighter bonds on $\mathsf{R}_{\rho,\tilde{\rho}}(D_\rho, D_{\tilde{\rho}})$ for the important class of difference distortion measures.

**Theorem 6.** *If*

(i) $\rho, \tilde{\rho}$ *are continuous*

(ii) *there exists* $a \geq 0$, $c \geq 0$, $\varepsilon > 0$ *such that for all* $(x, y) \in A \triangleq \{(x, y) : \rho(x, y) > a\}$

$$\sup_{z:\|z\|_\infty \leq \varepsilon} \rho(x, y + z) \leq c\rho(x, y)$$

(iii) $P \ll \lambda_{\mathbb{R}^m}$

*then for all* $\delta > 0$

$$\inf_{Q\in\mathcal{Q}_2(D_\rho+\delta, D_{\tilde{\rho}}-\delta)} I(Q) \leq \inf_{Q\in\mathcal{Q}_1(D_\rho, D_{\tilde{\rho}})} I(Q) \leq \inf_{Q\in\mathcal{Q}_2(D_\rho, D_{\tilde{\rho}})} I(Q).$$

*If, in addition,*

(iv) $\inf_{Q\in\mathcal{Q}_2(D_\rho, D_{\tilde{\rho}})} I(Q)$ *is continuous at* $(D_\rho, D_{\tilde{\rho}})$ *(as a function of* $(D_\rho, D_{\tilde{\rho}})$*)*

*then*

$$\inf_{Q\in\mathcal{Q}_1(D_\rho, D_{\tilde{\rho}})} I(Q) = \inf_{Q\in\mathcal{Q}_2(D_\rho, D_{\tilde{\rho}})} I(Q).$$

We say that $\rho$ and $\tilde{\rho}$ are *difference distortion measures* if $\rho(x, y)$ and $\tilde{\rho}(x, y)$ are functions of $y - x$. With some abuse of notation we shall write $\rho(y-x)$ and $\tilde{\rho}(y-x)$ in this case. The next theorem provides a lower bound on $\mathsf{R}_{\rho,\tilde{\rho}}(D_\rho, D_{\tilde{\rho}})$, similar to the Shannon lower bound for $\mathsf{R}_\rho(D_\rho)$.

**Theorem 7.** *Let* $\rho, \tilde{\rho}$ *be difference distortion measures, and let* $P \ll \lambda_{\mathbb{R}^m}$ *have finite differential entropy. If there exist* $\eta, \tilde{\eta} \geq 0$, *and* $\alpha$, *such that* $f : \mathbb{R}^m \to \mathbb{R}_+$ *defined by*

$$f(z) \triangleq \exp\big(-\alpha - \eta\rho(z) + \tilde{\eta}\tilde{\rho}(z)\big)$$

*satisfies*

$$\int f(z)dz = 1,$$
$$\int \rho(z)f(z)dz = D_\rho,$$
$$\int \tilde{\rho}(z)f(z)dz = D_{\tilde{\rho}},$$

*then*

$$\inf_{Q\in\mathcal{Q}_2(D_\rho, D_{\tilde{\rho}})} I(Q) \geq \max\big\{0, h(X) - h(Z)\big\} = \max\big\{0, h(X) - \alpha - \eta D_\rho + \tilde{\eta} D_{\tilde{\rho}}\big\}, \tag{9}$$

*where* $X \sim P$ *and* $Z$ *has density* $f$. *If, in addition, there exists a random variable* $Y$ *independent of* $Z$ *such that* $X = Y + Z$, *then we have equality in* (9).

**Example 4.** Let $\mathcal{X} = \mathcal{Y} = \mathbb{R}^2$,

$$\rho(\boldsymbol{x}, \boldsymbol{y}) = (\boldsymbol{y} - \boldsymbol{x})^T \begin{pmatrix} 1 & 0 \\ 0 & 1 \end{pmatrix} (\boldsymbol{y} - \boldsymbol{x}),$$
$$\tilde{\rho}(\boldsymbol{x}, \boldsymbol{y}) = (\boldsymbol{y} - \boldsymbol{x})^T \begin{pmatrix} a & 0 \\ 0 & b \end{pmatrix} (\boldsymbol{y} - \boldsymbol{x}),$$

with $a \geq b > 0$, and let $X$ be Gaussian with mean $\mathbf{0}$ and covariance matrix $\boldsymbol{I}$. The asymptotic expression (and upper bound) given by Theorem 3 is

$$\mathsf{D}_{\rho,\tilde{\rho}}(\infty, D_\rho) = aD_\rho \tag{10}$$

and on the boundary

$$\mathsf{D}_{\rho,\tilde{\rho}}(\mathsf{R}_\rho(D_\rho), D_\rho) = \frac{1}{2}(a+b)D_\rho. \tag{11}$$

We now apply Theorems 6 and 7 to compute $\mathsf{D}_{\rho,\tilde{\rho}}(R, D_\rho)$ for intermediate values of $R$. The density of $Z$ from Theorem 7 is given by

$$f(z) = \exp\left(\alpha - \boldsymbol{z}^T \begin{pmatrix} \eta - a\tilde{\eta} & 0 \\ 0 & \eta - b\tilde{\eta} \end{pmatrix} \boldsymbol{z}\right).$$

Let

$$\begin{aligned} \sigma^2 &\triangleq 1/(2\eta - 2a\tilde{\eta}), \\ \tilde{\sigma}^2 &\triangleq 1/(2\eta - 2b\tilde{\eta}), \end{aligned}$$

and note that $0 < \tilde{\sigma}^2 \leq \sigma^2$. With this, $f$ is a Gaussian density, with two independent components with mean zero and variances $\sigma^2$ and $\tilde{\sigma}^2$. For the bound on $\inf_{Q \in \mathcal{Q}_2(D_\rho, D_{\tilde{\rho}})} I(Q)$ given by Theorem 7 to be tight, we need to show that $X = Y + Z$ for some independent random variable $Y$. This is the case if we need $\sigma^2 \leq 1$ (and hence also $\tilde{\sigma}^2 \leq 1$).

In terms of $\sigma^2$ and $\tilde{\sigma}^2$, we have

$$\begin{aligned} \mathbb{E}\rho(Z) &= \sigma^2 + \tilde{\sigma}^2, \\ \mathbb{E}\tilde{\rho}(Z) &= a\sigma^2 + b\tilde{\sigma}^2, \\ h(X) - h(Z) &= -\frac{1}{2}\log(\sigma^2) - \frac{1}{2}\log(\tilde{\sigma}^2). \end{aligned}$$

A short computation reveals that for

$$\begin{aligned} \sigma^2 &= \frac{1}{2}\big(1 + \sqrt{1 - \exp(-2r)}\big)D_\rho, \\ \tilde{\sigma}^2 &= \frac{1}{2}\big(1 - \sqrt{1 - \exp(-2r)}\big)D_\rho, \end{aligned}$$

we have

$$\begin{aligned} \mathbb{E}\rho(Z) &= D_\rho, \\ \mathbb{E}\tilde{\rho}(Z) &= \frac{1}{2}\big((a+b) + \sqrt{1 - \exp(-2r)}(a-b)\big)D_\rho, \\ h(X) - h(Z) &= \mathsf{R}_\rho(D_\rho) + r. \end{aligned}$$

Thus, by Theorems 6 and 7,

$$\mathsf{D}_{\rho,\tilde{\rho}}(\mathsf{R}_\rho(D_\rho) + r, D_\rho) \leq \frac{1}{2}\big((a+b) + \sqrt{1 - \exp(-2r)}(a-b)\big)D_\rho.$$

And for

$$0 < D_\rho \leq 2/\big(1 + \sqrt{1 - \exp(-2r)}\big),$$

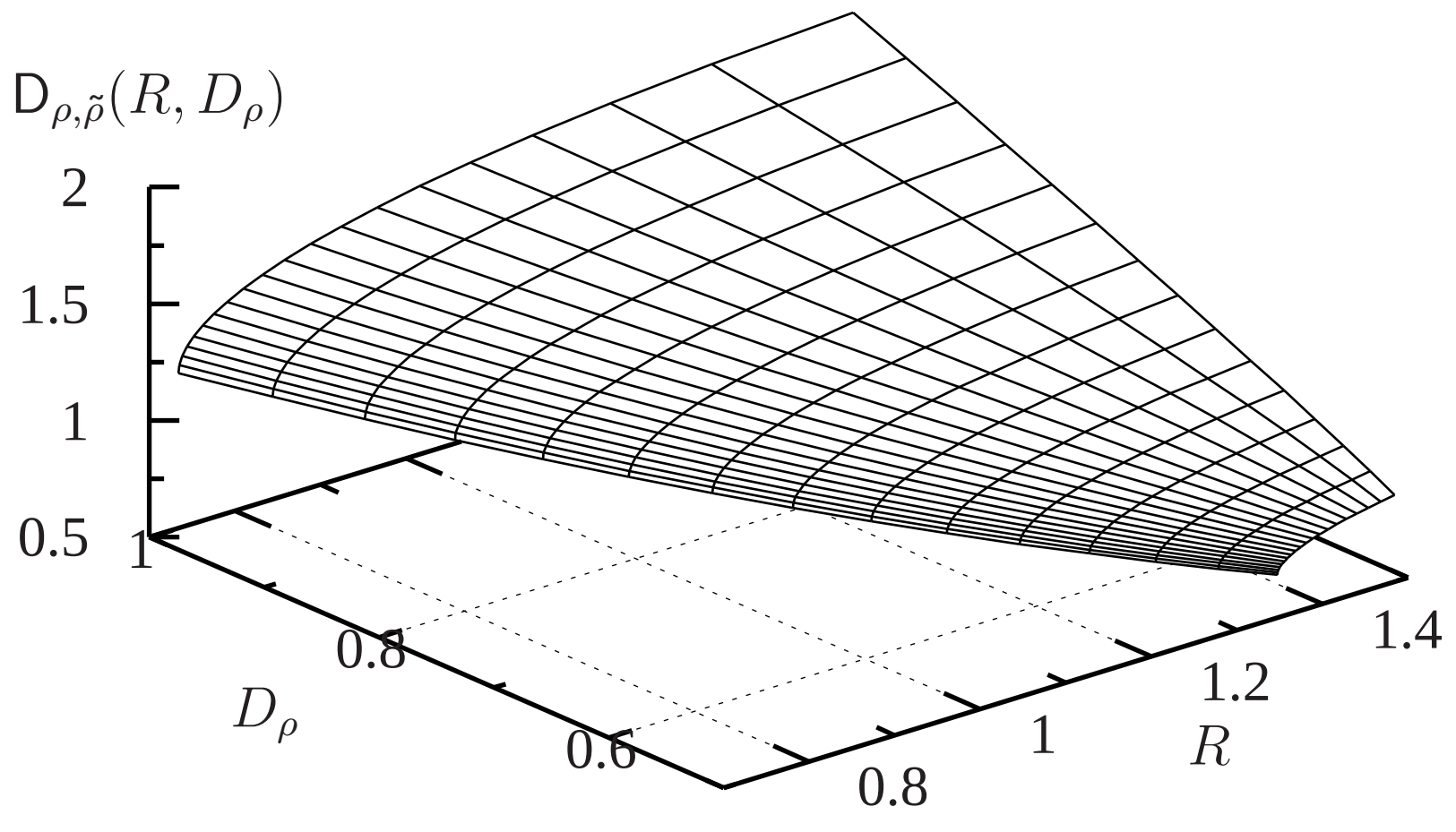

Fig. 3. $\mathsf{D}_{\rho,\tilde{\rho}}(R, D_\rho)$ from Example 4 with $a = 2$ and $b = 0.5$.

we have $\sigma^2 \leq 1$ and hence this bound is tight. In particular, this is the case for $0 < D_\rho \leq 1$. As a quick sanity check, we see that indeed

$$\lim_{r\to 0} \mathsf{D}_{\rho,\tilde{\rho}}(\mathsf{R}_\rho(D_\rho) + r, D_\rho) = \frac{1}{2}(a+b)D_\rho,$$
$$\lim_{r\to \infty} \mathsf{D}_{\rho,\tilde{\rho}}(\mathsf{R}_\rho(D_\rho) + r, D_\rho) = aD_\rho,$$

which are the values found in (10) and (11).

For $0 < D_\rho \leq 1$, the ratio between the limiting expression as $r \to \infty$ and the value for finite $r$ is independent of $D_\rho$ and given by

$$\mathsf{D}_{\rho,\tilde{\rho}}(\mathsf{R}_\rho(D_\rho) + r, D_\rho)/\mathsf{D}_{\rho,\tilde{\rho}}(\infty, D_\rho) = \big((a+b) + \sqrt{1-\exp(-2r)}(a-b)\big)/2a.$$

We see that this converges to one quickly as $r \to \infty$, as is shown in Figure 4. Hence in this case the limiting expression found in Theorem 3 is approached rapidly, and is hence a fairly tight upper bound on $\mathsf{D}_{\rho,\tilde{\rho}}(\mathsf{R}_\rho(D_\rho) + r, D_\rho)$ even for small values of $r$.

$\Diamond$

### *D. Choosing a "Representative" of a Class of Distortion Measures*

Let $\Gamma$ and $\widetilde{\Gamma}$ denote classes of distortion measures. In this section, we consider the question of how a good "representative" $\rho \in \Gamma$ of $\widetilde{\Gamma}$ can be chosen (in a sense to be made precise).

Consider again the oracle producing source codes as mentioned in the introduction, but assume this time that when queried, we can also supply the oracle with a distortion measure $\rho \in \Gamma$. The oracle then produces a source code $f_n$ such that

$$\frac{1}{n}\log|f_n(\mathcal{X}^n)| \leq R$$
$$\mathbb{E}\rho_n(X^n, f_n(X^n)) \leq \mathsf{D}_\rho(R) + \Delta_\rho.$$

Knowing the set of all $\{\Delta_\rho\}_{\rho\in\Gamma}$, and given a $\widetilde{\Gamma}$, how should we choose $\rho \in \Gamma$ to query the oracle with such that $f^n$ will "work well" for all $\tilde{\rho} \in \widetilde{\Gamma}$?

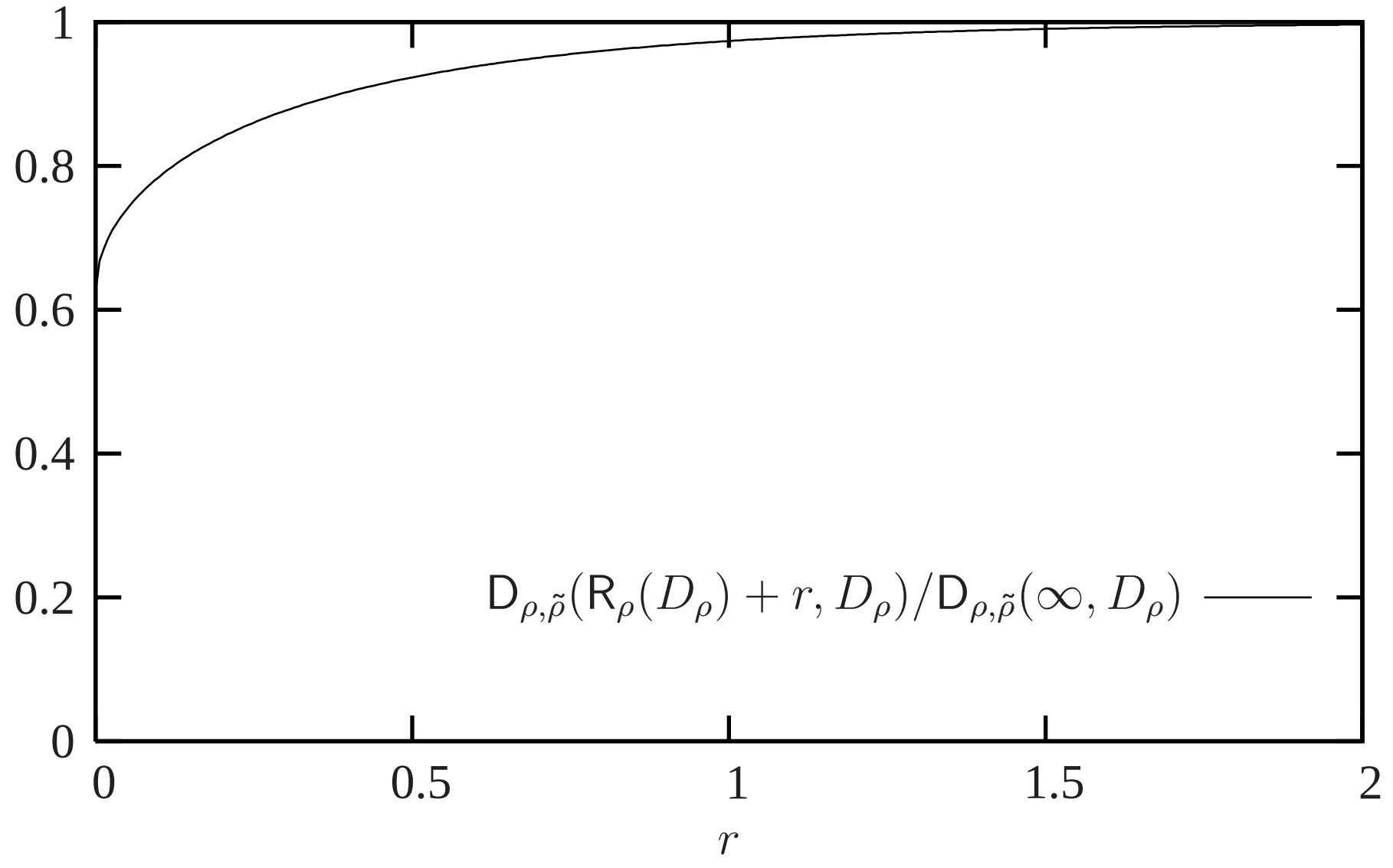

Fig. 4. $\mathsf{D}_{\rho,\tilde{\rho}}(\mathsf{R}_\rho(D_\rho)+r, D_\rho)/\mathsf{D}_{\rho,\tilde{\rho}}(\infty, D_\rho)$ from Example 4 as a function of $r$ with $a = 2$, $b = 0.5$, for all values $0 < D_\rho \leq 1$. Note that for an excess rate of $r = 0.5$, we are already at over $90\%$ of the limiting value, at excess rate of $r = 1$, we are at over $97\%$ of the limiting value.

This problem has the following operational significance. Assume we have a collection $\Gamma$ of tractable distortion measures (i.e., distortion measures for which we are able to design good source codes). Assume furthermore, we know that the true distortion measure lies in some class $\widetilde{\Gamma}$. We can choose a source code designed for one of the tractable distortion measures in $\Gamma$. We are then using this source code with respect to any of the distortion measures in $\widetilde{\Gamma}$. While in the previous sections we were only analyzing the performance guarantees under mismatched distortion measures, here we also get to choose $\rho \in \Gamma$ in order to minimize the mismatch.

The parameters $\{\Delta_\rho\}_{\rho\in\Gamma}$ allow to account for the difficulty of constructing a source code for distortion measure $\rho$ (see also Example 5 below). Note, however, that there are several reasonable ways in which "work well" in the last paragraph can be defined. We will consider two such definitions in the following. For rate $R$, define

$$\mathsf{D}_{\Gamma,\widetilde{\Gamma}}\big(R, \{\Delta_\rho\}\big) \triangleq \inf_{\rho\in\Gamma} \sup_{\tilde{\rho}\in\widetilde{\Gamma}} \mathsf{D}_{\rho,\tilde{\rho}}(R, \mathsf{D}_\rho(R)+\Delta_\rho),$$

$$\mathsf{\Delta}_{\Gamma,\widetilde{\Gamma}}\big(R, \{\Delta_\rho\}\big) \triangleq \inf_{\rho\in\Gamma} \sup_{\tilde{\rho}\in\widetilde{\Gamma}} \big(\mathsf{D}_{\rho,\tilde{\rho}}(R, \mathsf{D}_\rho(R)+\Delta_\rho) - \mathsf{D}_{\tilde{\rho}}(R)\big).$$

We assume throughout that the $\{\Delta_\rho\}_{\rho\in\Gamma}$ satisfy

$$\inf_{\rho\in\Gamma} \Delta_\rho > 0.$$

The next example illustrates why introducing $\{\Delta_\rho\}_{\rho\in\Gamma}$ is necessary.

**Example 5.** Fix distortion measures $\rho, \tilde{\rho}$, and let $\Gamma \triangleq \{a\rho\}_{a\geq 1}$. All distortion measures in $\Gamma$ are equivalent (in the sense that constructing source codes for $\rho$ is as difficult as constructing source codes for any $a\rho$). So we should have that all $a\rho$ represent $\tilde{\rho}$ equally well (in the sense that for appropriately chosen $D_{a\rho}$, $\mathsf{D}_{a\rho,\tilde{\rho}}(R, D_{a\rho})$ is the same for all $a \geq 1$). As we will see in a moment, this imposes the introduction of the quantity $\{\Delta_{a\rho}\}_{a\geq 1}$.

For any *fixed* $D_\rho$, we have

$$\mathsf{D}_{a\rho,\tilde{\rho}}(R, D_\rho) = \mathsf{D}_{\rho,\tilde{\rho}}(R, D_\rho/a)$$

which goes either to $0$ (if $R \geq \mathsf{R}_\rho(0)$) or to $-\infty$ as $a \to \infty$. This shows that we should look at source codes constructed with distortion level relative to $\mathsf{D}_{a\rho}(R)$ Assume then we try to minimize $\mathsf{D}_{a\rho,\tilde{\rho}}(R, \mathsf{D}_{a\rho}(R)+\Delta)$ for some *fixed* $\Delta > 0$. We have

$$\mathsf{D}_{a\rho,\tilde{\rho}}(R, \mathsf{D}_{a\rho}(R) + \Delta) = \mathsf{D}_{\rho,\tilde{\rho}}(R, \mathsf{D}_\rho(R) + \Delta/a).$$

Thus, again, the minimum is achieved as $a \to \infty$, irrespective of the choice of $\tilde{\rho}$. This shows, that we should not choose $\Delta_{a\rho}$ as a constant. The natural choice in this example is $\Delta_{a\rho} = a\Delta$, for which

$$\mathsf{D}_{a\rho,\tilde{\rho}}(R, \mathsf{D}_{a\rho}(R) + \Delta_{a\rho}) = \mathsf{D}_{\rho,\tilde{\rho}}(R, \mathsf{D}_\rho(R) + \Delta),$$

as expected. $\Diamond$

The following two corollaries of Theorem 1 and 2, respectively, establish the operational meaning of $\mathsf{D}_{\Gamma,\widetilde{\Gamma}}\big(R, \{\Delta_\rho\}\big)$ and $\mathsf{\Delta}_{\Gamma,\widetilde{\Gamma}}\big(R, \{\Delta_\rho\}\big)$.

**Corollary 8.** *Let $\Gamma, \widetilde{\Gamma}$ be classes of distortion measures such that for all $\rho \in \Gamma$ there exists a $y_0 = y_0(\rho) \in \mathcal{Y}$ satisfying $\mathbb{E}_P \rho(X, y_0) < \infty$. For every $\rho \in \Gamma$, $R > 0$, and $D_{\widetilde{\Gamma}}, \Delta_{\widetilde{\Gamma}}$ such that*

$$\begin{aligned} 0 \leq & D_{\widetilde{\Gamma}} < \lim_{\delta \downarrow 0} \mathsf{D}_{\Gamma,\widetilde{\Gamma}}\big(R, \{\Delta_\rho - \delta\}\big), \\ 0 \leq & \Delta_{\widetilde{\Gamma}} < \lim_{\delta \downarrow 0} \mathsf{\Delta}_{\Gamma,\widetilde{\Gamma}}\big(R, \{\Delta_\rho - \delta\}\big), \end{aligned}$$

a) *there exists $\tilde{\rho} \in \widetilde{\Gamma}$ and sequences of source codes $\{f_n\}_{n \geq 1}$ such that*

$$\begin{aligned} \lim_{n\to\infty} \frac{1}{n} \log |f_n(\mathcal{X}^n)| &\leq R, \\ \limsup_{n\to\infty} \mathbb{E}\rho_n(X^n, f_n(X^n)) &\leq \mathsf{D}_\rho(R) + \Delta_\rho, \\ \liminf_{n\to\infty} \mathbb{E}\tilde{\rho}_n(X^n, f_n(X^n)) &\geq D_{\widetilde{\Gamma}}. \end{aligned}$$

b) *there exists $\tilde{\rho} \in \widetilde{\Gamma}$ and sequences of source codes $\{f_n\}_{n \geq 1}$ such that*

$$\begin{aligned} \lim_{n\to\infty} \frac{1}{n} \log |f_n(\mathcal{X}^n)| &\leq R, \\ \limsup_{n\to\infty} \mathbb{E}\rho_n(X^n, f_n(X^n)) &\leq \mathsf{D}_\rho(R) + \Delta_\rho, \\ \liminf_{n\to\infty} \big(\mathbb{E}\tilde{\rho}_n(X^n, f_n(X^n)) - \mathsf{D}_{\tilde{\rho}}(R)\big) &\geq \Delta_{\widetilde{\Gamma}}. \end{aligned}$$

**Corollary 9.** a) *For every $\delta > 0$ there exists $\rho \in \Gamma$ such that if $f_n : \mathcal{X}^n \to \mathcal{Y}^n$ satisfies*

$$\begin{aligned} \frac{1}{n} \log |f_n(\mathcal{X}^n)| &= R, \\ \mathbb{E}\rho_n(X^n, f_n(X^n)) &\leq \mathsf{D}_\rho(R) + \Delta_\rho, \end{aligned}$$

*then*

$$\sup_{\tilde{\rho} \in \widetilde{\Gamma}} \mathbb{E}\tilde{\rho}_n(X^n, f_n(X^n)) \leq \mathsf{D}_{\Gamma,\widetilde{\Gamma}}\big(R, \{\Delta_\rho\}\big) + \delta.$$

b) *For every $\delta > 0$ there exists $\rho \in \Gamma$ such that if $f_n : \mathcal{X}^n \to \mathcal{Y}^n$ satisfies*

$$\begin{aligned} \frac{1}{n} \log |f_n(\mathcal{X}^n)| &= R, \\ \mathbb{E}\rho_n(X^n, f_n(X^n)) &\leq \mathsf{D}_\rho(R) + \Delta_\rho, \end{aligned}$$

*then*

$$\sup_{\tilde{\rho} \in \Gamma} \big(\mathbb{E}\tilde{\rho}_n(X^n, f_n(X^n)) - \mathsf{D}_{\tilde{\rho}}(R)\big) \leq \mathsf{\Delta}_{\Gamma,\widetilde{\Gamma}}\big(R, \{\Delta_\rho\}\big) + \delta.$$

Corollaries 8 and 9 allow us to make guarantees about the performance of a source codes constructed with respect to the best "representative" $\rho \in \Gamma$ of $\widetilde{\Gamma}$. Indeed, by Corollary 9, there exists $\rho \in \Gamma$ such that if $f_n : \mathcal{X}^n \to \mathcal{Y}^n$ is a source code of rate $R$ designed for distortion measure $\rho$ and distortion level $D_\rho(R) + \Delta_\rho$, then $f_n$ is also a source code for any distortion measure $\tilde{\rho} \in \widetilde{\Gamma}$ and distortion level $\mathsf{D}_{\Gamma,\widetilde{\Gamma}}\big(R, \{\Delta_\rho\}\big) + \delta$. Moreover, this is essentially the best guarantee one can make, since by Corollary 8 there exist source codes with same blocklength $n$ and same rate $R$ designed for any distortion measure $\rho \in \Gamma$ and distortion level $D_\rho(R) + \Delta_\rho$ that result in a distortion level of more than

$$\mathsf{D}_{\Gamma,\widetilde{\Gamma}}\big(R - \delta(n), \{\Delta_\rho - \tilde{\delta}(n)\}\big) - \delta(n)$$

for some distortion measure $\tilde{\rho} \in \widetilde{\Gamma}$, and with $\delta(n), \tilde{\delta}(n) \to 0$ as $n \to \infty$.

**Example 6.** Let $\widetilde{\Gamma} = \{\tilde{\rho}\}$, and

$$\Gamma \triangleq \big\{\rho(\boldsymbol{x}, \boldsymbol{y}) = w_{\boldsymbol{x}} \left\|\boldsymbol{y} - \boldsymbol{x}\right\|_2^2 : w \in \mathcal{W} \subset (\mathcal{X} \to \mathbb{R}_+)\big\},$$

Let $P \in \mathcal{P}(\mathcal{X} \times \mathcal{Y})$ be such that $\mathbb{E}_P w_{\boldsymbol{X}} \left\|\boldsymbol{X}\right\|_2^2 < \infty$ for all $w \in \mathcal{W}$. In [13], the authors show how vector quantizers can be relatively easily constructed for distortion measures in the class $\Gamma$ defined here. Given a more sophisticated distortion measure $\tilde{\rho}$, it is thus of interest to find the "closest" $\rho \in \Gamma$ to $\tilde{\rho}$. In other words, for some $\delta > 0$, we want to find a $\rho \in \Gamma$ such that

$$\mathsf{D}_{\rho,\tilde{\rho}}(R, \mathsf{D}_\rho(R) + \Delta_\rho) \leq \mathsf{D}_{\Gamma,\widetilde{\Gamma}}(R) + \delta.$$

Computing $\mathsf{D}_{\Gamma,\widetilde{\Gamma}}\big(R, \{\Delta_\rho\}\big)$ could be done numerically; to obtain some insight we will instead minimize $\mathsf{D}_{\rho,\tilde{\rho}}(\infty, \mathsf{D}_\rho(R) + \Delta_\rho)$. This will lead to an upper bound on $\mathsf{D}_{\Gamma,\widetilde{\Gamma}}\big(R, \{\Delta_\rho\}\big)$, and thus allows us to make performance guarantees. Moreover, as we have seen in Example 4, this bound can be quite good even for finite values of $R$. To be specific, let $\tilde{\rho}(\boldsymbol{x}, \boldsymbol{y}) = (\boldsymbol{y} - \boldsymbol{x})^T \widetilde{\boldsymbol{W}}_{\boldsymbol{x}} (\boldsymbol{y} - \boldsymbol{x})$ for $\widetilde{\boldsymbol{W}}_{\boldsymbol{x}}$ positive definite $P$ almost everywhere. Let $w^\rho \in \mathcal{W}$ be the weight function corresponding to distortion measure $\rho \in \Gamma$. Then from Example 1,

$$\begin{aligned}
&\mathsf{D}_{\rho,\tilde{\rho}}(\infty, \mathsf{D}_\rho(R) + \Delta_\rho) \\
&\qquad = (\mathsf{D}_\rho(R) + \Delta_\rho) \min\big\{\eta : \widetilde{\boldsymbol{W}}_{\boldsymbol{x}} - \eta w^\rho_{\boldsymbol{x}} \boldsymbol{I} \leq 0 \ P \text{ a.e}\big\} \\
&\qquad = (\mathsf{D}_\rho(R) + \Delta_\rho) \operatorname*{ess\,sup}_{x \in \mathcal{X}} \lambda_1(\widetilde{\boldsymbol{W}}_{\boldsymbol{x}}) / w^\rho_{\boldsymbol{x}},
\end{aligned}$$

where $\lambda_1(\widetilde{\boldsymbol{W}}_{\boldsymbol{x}})$ is the largest eigenvalue of $\widetilde{\boldsymbol{W}}_{\boldsymbol{x}}$, and where the essential supremum is with respect to $P$. Hence

$$\mathsf{D}_{\Gamma,\widetilde{\Gamma}}\big(R, \{\Delta_\rho\}\big) \leq \inf_{\rho \in \Gamma} \Big((\mathsf{D}_\rho(R) + \Delta_\rho) \operatorname*{ess\,sup}_{x \in \mathcal{X}} \lambda_1(\widetilde{\boldsymbol{W}}_{\boldsymbol{x}}) / w^\rho_{\boldsymbol{x}}\Big).$$

In other words, the optimal "representative" $\rho \in \Gamma$ of $\tilde{\rho}$ finds the best tradeoff between the difficulty of constructing source codes for $\rho$ (captured by the term $\mathsf{D}_\rho(R) + \Delta_\rho$) and the closeness to $\tilde{\rho}$ (captured by the term $\operatorname*{ess\,sup}_{x \in \mathcal{X}} \lambda_1(\widetilde{\boldsymbol{W}}_{\boldsymbol{x}}) / w^\rho_{\boldsymbol{x}}$). $\Diamond$

In the last example, we have taken a sophisticated distortion measure $\tilde{\rho}$ and found a good tractable approximation in $\Gamma$ for it. This approach poses the following question. Even if $\tilde{\rho}$ is a very good model for (say) the human visual system, it will certainly be different from it. In this situation, it is not clear if minimizing $\mathsf{D}_{\rho,\tilde{\rho}}(R, \mathsf{D}_\rho(R) + \Delta_\rho)$ is meaningful. Indeed, if $\rho^*$ is the distortion measure implemented by the human visual system, we should really be minimizing $\mathsf{D}_{\rho,\rho^*}(R, \mathsf{D}_\rho(R) + \Delta_\rho)$ instead. The next theorem provides conditions under which $\mathsf{D}_{\rho,\tilde{\rho}}(R, \mathsf{D}_\rho(R) + \Delta_\rho)$ and $\mathsf{D}_{\rho,\rho^*}(R, \mathsf{D}_\rho(R) + \Delta_\rho)$ are close and hence the approach of Example 6 is reasonable.

**Proposition 10.** *Let $\rho_1, \rho_2, \rho_3$ be continuous distortion measures. Then*

$$\mathsf{D}_{\rho_1,\rho_3}(R, D) \leq \mathsf{D}_{\rho_1,\rho_2}(R, D) + \mathbb{E}_P(\sup_{y \in \mathcal{Y}} \rho_3(X, y) - \rho_2(X, y))$$

*and*

$$\mathsf{D}_{\rho_1,\rho_3}(R,D) \geq \mathsf{D}_{\rho_1,\rho_2}(R,D) - \mathbb{E}_P \sup_{y\in\mathcal{Y}} |\rho_3(X,y)-\rho_2(X,y)|.$$

**Example 7.** Setting $\rho_1=\rho_2$, Proposition 10 shows that

$$\left|\mathsf{D}_{\rho_2,\rho_3}(R,D_{\rho_1}) - D_{\rho_2}(R)\right| \leq \mathbb{E}_P \sup_{y\in\mathcal{Y}} |\rho_3(X,y)-\rho_2(X,y)|.$$

Thus if

$$\mathbb{E}_P \sup_{y\in\mathcal{Y}} |\rho_3(X,y)-\rho_2(X,y)|$$

is small then the distortion measures $\rho_2$ and $\rho_3$ are almost equivalent (from the point of source coding). Moreover, if $\rho_3$ is the actual distortion measure (implemented, e.g., by the human visual system), and $\rho_2$ is a sophisticated model for it (e.g., $\rho_2(\boldsymbol{x},\boldsymbol{y}) = (\boldsymbol{y}-\boldsymbol{x})^T\widetilde{\boldsymbol{W}_{\boldsymbol{x}}}(\boldsymbol{y}-\boldsymbol{x})$ as in Example 6), then small

$$\mathbb{E}_P \sup_{y\in\mathcal{Y}} |\rho_3(X,y)-\rho_2(X,y)|$$

guarantees that minimizing $\mathsf{D}_{\rho_1,\rho_2}(R,D_{\rho_1}+\Delta_{\rho_1})$ over all $\rho_1\in\Gamma$ (as is done in Example 6) is essentially equivalent to minimizing $\mathsf{D}_{\rho_1,\rho_3}(R,D_{\rho_1}+\Delta_{\rho_1})$. Hence, when constructing a model $\rho_2$ for the distortion measure $\rho_3$ implemented by the human visual system, it is reasonable to choose the model parameters such that

$$\mathbb{E}_P \sup_{y\in\mathcal{Y}} |\rho_3(X,y)-\rho_2(X,y)|$$

is minimized. $\Diamond$

## III. Proofs

### A. Proof of Theorem 1

A slight modification of Lemma 9.3.1 and the first part of the proof of Theorem 9.6.2 in [14] show that for every $\delta>0$ there exists a sequence of source codes $\{\tilde{f}_n\}_{n\geq 1}$ such that

$$\begin{gathered}\lim_{n\to\infty} P^n(A_n) = 0,\\ \lim_{n\to\infty}\frac{1}{n}\log|\tilde{f}_n(\mathcal{X}^n)| \leq R,\end{gathered} \tag{12}$$

where

$$A_n \triangleq \{x^n : \rho_n(x^n,\tilde{f}_n(x^n)) > D_\rho - \delta/2\} \cup \{x^n : \tilde{\rho}_n(x^n,\tilde{f}_n(x^n)) < D_{\tilde{\rho}}\}.$$

Let

$$B_n \triangleq \{x^n : \rho_n(x^n,\tilde{f}^n(x^n)) > D_\rho - \delta/2\} \subset A_n,$$

and set

$$f_n(x^n) \triangleq \begin{cases}\boldsymbol{y}_0 & \text{if } x^n\in B_n,\\ \tilde{f}_n(x^n) & \text{else,}\end{cases}$$

where $\boldsymbol{y}_0 \triangleq (y_0,\ldots,y_0)\in\mathcal{Y}^n$. We have $|f_n(\mathcal{X}^n)| \leq |\tilde{f}_n(\mathcal{X}^n)|+1$ and hence by (12)

$$\lim_{n\to\infty}\frac{1}{n}\log|f^n(\mathcal{X}^n)| = \lim_{n\to\infty}\frac{1}{n}\log|\tilde{f}^n(\mathcal{X}^n)| \leq R.$$

Moreover,

$$\mathbb{E}\rho_n(X^n,f_n(X^n)) \leq D_\rho - \delta/2 + \mathbb{E}(\rho_n(X^n,f_n(X^n));B_n),$$

and for any $b \geq 0$

$$\begin{aligned}\mathbb{E}(\rho_n(X^n, & f_n(X^n)); B_n) \\ &= \frac{1}{n}\sum_{i=1}^{n}\mathbb{E}(\rho(X_i, y_0); B_n) \\ &\leq \frac{1}{n}\sum_{i=1}^{n}\Big(\mathbb{E}\big(\rho(X_i, y_0); \{\rho(X_i, y_0) \leq b\} \cap B_n\big) + \mathbb{E}\big(\rho(X_i, y_0); \{\rho(X_i, y_0) > b\}\big)\Big) \\ &\leq bP^n(A_n) + \mathbb{E}_P\big(\rho(X, y_0); \{\rho(X, y_0) > b\}\big).\end{aligned}$$

Since $\mathbb{E}_P\rho(X, y_0) < \infty$, there exists $b > 0$ such that $\mathbb{E}_P\big(\rho(X, y_0); \{\rho(X, y_0) > b\}\big) \leq \delta/2$. Hence using (12),

$$\limsup_{n\to\infty}\mathbb{E}\rho_n(X^n, f_n(X^n)) \leq D_\rho.$$

Finally,

$$\begin{aligned}\mathbb{E}\tilde{\rho}(X^n, f_n(X^n)) &\geq \mathbb{E}(\tilde{\rho}(X^n, f_n(X^n)); A_n^c) \\ &\geq D_{\tilde{\rho}}P^n(A_n^c),\end{aligned}$$

and hence by (12)

$$\liminf_{n\to\infty}\mathbb{E}\tilde{\rho}_n(X^n, f_n(X^n)) \geq D_{\tilde{\rho}}.$$

*B. Proof of Theorem 2*

Let $\tilde{\rho}' \triangleq -\tilde{\rho}$. If

$$\begin{aligned}\frac{1}{n}\log|f_n(\mathcal{X}^n)| &= R, \\ \mathbb{E}\rho_n(X^n, f_n(X^n)) &\leq D_\rho, \\ \mathbb{E}\tilde{\rho}_n(X^n, f_n(X^n)) &\geq D_{\tilde{\rho}},\end{aligned}$$

then we also have

$$\mathbb{E}\tilde{\rho}'_n(X^n, f_n(X^n)) \leq -D_{\tilde{\rho}} \triangleq D_{\tilde{\rho}'}.$$

By [5, Theorem 1.b], for every $\delta > 0$ there exists $Q \in \mathcal{P}(\mathcal{X} \times \mathcal{Y})$ such that $Q_{\mathcal{X}} = P$ and

$$\begin{aligned}I(Q) &\leq R + \delta, \\ \mathbb{E}_Q\rho &\leq D_\rho, \\ \mathbb{E}_Q\tilde{\rho}' &\leq D_{\tilde{\rho}'}.\end{aligned}$$

Therefore

$$D_{\tilde{\rho}} \leq \mathbb{E}_Q\tilde{\rho} \leq \mathsf{D}_{\rho,\tilde{\rho}}(R + \delta, D_\rho),$$

and maximizing over the choice of $D_{\tilde{\rho}}$ yields the first part of the theorem.

For the second part, we need to show that $\mathsf{D}_{\rho,\tilde{\rho}}(\cdot, D_\rho)$ is continuous for $R > \mathsf{R}_\rho(D_\rho)$. We first show that $\mathsf{D}_{\rho,\tilde{\rho}}(\cdot, D_\rho)$ is concave. Fix $\delta > 0$. Let $Q_1, Q_2 \in \mathcal{P}(\mathcal{X} \times \mathcal{Y})$, both with $\mathcal{X}$ marginal $P$, and such that $I(Q_i) \leq R_i$, $\mathbb{E}_{Q_i}\rho \leq D_\rho$, and $\mathbb{E}_{Q_i}\tilde{\rho} \geq \mathsf{D}_{\rho,\tilde{\rho}}(R_i, D_\rho) - \delta$ for $i \in \{1, 2\}$. Setting $Q \triangleq \alpha Q_1 + (1-\alpha)Q_2$, we have $\mathbb{E}_Q\rho \leq D_\rho$ and

$$\begin{aligned}\mathbb{E}_Q\tilde{\rho} &= \alpha\mathbb{E}_{Q_1}\tilde{\rho} + (1-\alpha)\mathbb{E}_{Q_2}\tilde{\rho} \\ &\geq \alpha\mathsf{D}_{\rho,\tilde{\rho}}(R_1, D_\rho) + (1-\alpha)\mathsf{D}_{\rho,\tilde{\rho}}(R_2, D_\rho) - \delta.\end{aligned}$$

Since mutual information is convex in the conditional distribution [15, Corollary 5.5.5],

$$\begin{aligned} I(Q) &\leq \alpha I(Q_1) + (1-\alpha) I(Q_2) \\ &\leq \alpha R_1 + (1-\alpha) R_2. \end{aligned}$$

Hence

$$\mathsf{D}_{\rho,\tilde{\rho}}(\alpha R_1 + (1-\alpha)R_2, D_\rho) \geq \alpha \mathsf{D}_{\rho,\tilde{\rho}}(R_1, D_\rho) + (1-\alpha)\mathsf{D}_{\rho,\tilde{\rho}}(R_2, D_\rho) - \delta.$$

Since $\delta > 0$ is arbitrary, this proves concavity of $\mathsf{D}_{\rho,\tilde{\rho}}(\cdot, D_\rho)$. Moreover $\mathsf{D}_{\rho,\tilde{\rho}}(\cdot, D_\rho)$ is increasing, and therefore this implies that it is right-continuous except for possibly at the point $\mathsf{R}_\rho(D_\rho)$. From this, the result follows.

### *C. Proof of Theorem 3*

We first show that

$$\lim_{R\to\infty} \mathsf{D}_{\rho,\tilde{\rho}}(R, D_\rho) = \mathsf{D}_{\rho,\tilde{\rho}}(\infty, D_\rho).$$

$\mathsf{D}_{\rho,\tilde{\rho}}(\infty, D_\rho) < \infty$ by Assumption (iv), and therefore there exists $Q \in \mathcal{P}(\mathcal{X} \times \mathcal{Y})$ such that $Q_{\mathcal{X}} = P$, $\mathbb{E}_Q \rho \leq D_\rho$, and $\mathbb{E}_Q \tilde{\rho} \geq \mathsf{D}_{\rho,\tilde{\rho}}(\infty, D_\rho) - \varepsilon$. Let $K_i \subset \mathcal{X} \times \mathcal{Y}$ be compact and such that $Q(K_i) \geq 1 - 1/i$ for all $i \geq 1$. Thus $Q(\cup_{i\geq 1} K_i) = 1$, and therefore by dominated convergence (using Assumption (iv) for the first line and Assumption (ii) for the second line)

$$\begin{aligned} \lim_{I\to\infty} \mathbb{E}_Q(\tilde{\rho}; \cup_{i=1}^I K_i) &= \mathbb{E}_Q \tilde{\rho}, \\ \lim_{I\to\infty} \mathbb{E}_P(\rho(X, y_0); (\cup_{i=1}^I K_i)^c) &= 0. \end{aligned}$$

Hence there exists a compact $K \subset \mathcal{X} \times \mathcal{Y}$ such that

$$\mathbb{E}_Q(\tilde{\rho}; K) \geq \mathbb{E}_Q \tilde{\rho} - \varepsilon \tag{13}$$
$$\mathbb{E}_P(\rho(X, y_0); K^c) \leq \varepsilon. \tag{14}$$

Since $\rho$ and $\tilde{\rho}$ are continuous by Assumption (i), they are uniformly continuous on the compact set $K$. Hence there exists $\delta > 0$ such that

$$\begin{aligned} |\rho(x,y) - \rho(\tilde{x}, \tilde{y})| &< \varepsilon \\ |\tilde{\rho}(x,y) - \tilde{\rho}(\tilde{x}, \tilde{y})| &< \varepsilon, \end{aligned}$$

whenever $\|x - \tilde{x}\| + \|y - \tilde{y}\| < \delta$. Now, since $K$ is compact, there exists some finite $L$ and $\{x_\ell, y_\ell\}_{\ell=1}^L \subset K$ and a finite measurable partition $\{A_\ell\}_{\ell=1}^L$ of $K$ such that $\|x - x_\ell\| + \|y - y_\ell\| < \delta$ for all $(x, y) \in A_\ell$ and for all $\ell \in \{1, \ldots, L\}$.

Define

$$\widetilde{Y} \triangleq \begin{cases} y_0 & \text{if } (X, Y) \in K^c, \\ y_\ell & \text{if } (X, Y) \in A_\ell. \end{cases}$$

Since $K$ and $\{A_\ell\}_{\ell=1}^L$ are measurable, $\widetilde{Y}$ is a random variable. Let $\widetilde{Q}$ be the distribution of $(X, \widetilde{Y})$ when $(X, Y) \sim Q$. Since $\widetilde{Y}$ takes on at most $L + 1$ values, we have

$$I(\widetilde{Q}) \leq \log(L + 1) < \infty.$$

Moreover,

$$\begin{aligned}
\mathbb{E}_{\widetilde{Q}}\rho &= \sum_{\ell=1}^{L} \mathbb{E}_{\widetilde{Q}}(\rho; A_\ell) + \mathbb{E}_{\widetilde{Q}}(\rho; K^c) \\
&= \sum_{\ell=1}^{L} \mathbb{E}_P(\rho(X, y_\ell); A_\ell) + \mathbb{E}_P(\rho(X, y_0); K^c) \\
&\le \sum_{\ell=1}^{L} \mathbb{E}_Q(\rho(X, Y) + \varepsilon; A_\ell) + \varepsilon \\
&\le \mathbb{E}_Q\rho + 2\varepsilon \\
&\le D_\rho + 2\varepsilon,
\end{aligned}$$

where the first inequality follows from the uniform continuity of $\rho$ on $K$, and from (14). And

$$\begin{aligned}
\mathbb{E}_{\widetilde{Q}}\tilde{\rho} &\ge \sum_{\ell=1}^{L} \mathbb{E}_{\widetilde{Q}}(\tilde{\rho}; A_\ell) \\
&= \sum_{\ell=1}^{L} \mathbb{E}_P(\tilde{\rho}(X, y_\ell); A_\ell) \\
&\ge \sum_{\ell=1}^{L} \mathbb{E}_Q(\tilde{\rho}(X, Y) - \varepsilon; A_\ell) \\
&= \mathbb{E}_Q(\tilde{\rho}; K) - \varepsilon \\
&\ge \mathbb{E}_Q\tilde{\rho} - 2\varepsilon \\
&\ge \mathsf{D}_{\rho,\tilde{\rho}}(\infty, D_\rho) - 3\varepsilon,
\end{aligned}$$

where the second inequality follows from the uniform continuity of $\tilde{\rho}$ on $K$, and the third inequality form (13). Therefore

$$\begin{aligned}
\mathsf{D}_{\rho,\tilde{\rho}}(\infty, D_\rho) &\le \lim_{R\to\infty} \mathsf{D}_{\rho,\tilde{\rho}}(R, D_\rho + 2\varepsilon) + 3\varepsilon \\
&\le \mathsf{D}_{\rho,\tilde{\rho}}(\infty, D_\rho + 2\varepsilon) + 3\varepsilon.
\end{aligned}$$

Since $\mathsf{D}_{\rho,\tilde{\rho}}(\infty, \cdot)$ is concave, it is continuous at $D_\rho > \mathsf{D}_\rho(\infty)$ (Assumption (iii)). Hence taking the limit as $\varepsilon \to 0$ yields

$$\lim_{R\to\infty} \mathsf{D}_{\rho,\tilde{\rho}}(R, D_\rho) = \mathsf{D}_{\rho,\tilde{\rho}}(\infty, D_\rho).$$

We now show that

$$\mathbb{E}_P \sup_{y\in\mathcal{Y}} (\tilde{\rho}(X, y) - \eta\rho(X, y)) \tag{15}$$

is well defined. Let $\eta > 0$,

$$\begin{aligned}
f(x, y) &\triangleq \tilde{\rho}(x, y) - \eta\rho(x, y), \\
g(x) &\triangleq \sup_{y\in\mathcal{Y}} f(x, y).
\end{aligned}$$

By Assumption (i), $f$ is continuous, and hence

$$\sup_{y\in\mathbb{R}^m} f(x, y) = \sup_{y\in\mathbb{Q}^m} f(x, y).$$

As $\mathbb{Q}^m$ is countable, this last supremum is measurable, and hence $g$ is a measurable function. Moreover,

$$g(x) \ge f(x, y_0) \ge -\eta\rho(x, y_0),$$

and hence by Assumption (ii)

$$\mathbb{E}_P g^- < \infty,$$

where $g^- \triangleq \max\{0, -g\}$ is the negative part of $g$. Thus the expectation $\mathbb{E}_P g$ in (15) is well defined.

We next show that

$$\mathsf{D}_{\rho,\tilde{\rho}}(\infty, D_\rho) \leq \min_{\eta \geq 0} \eta D_\rho + \mathbb{E}_P g. \tag{16}$$

Consider

$$\mathsf{D}_{\rho,\tilde{\rho}}(\infty, D_\rho) = \sup_{\substack{Q \in \mathcal{P}(\mathcal{X}\times\mathcal{Y}): \\ Q_{\mathcal{X}} = P, \mathbb{E}_Q \rho \leq D_\rho}} \mathbb{E}_Q \tilde{\rho}.$$

The right hand side is linear in $Q$ with linear constraints. Since $D_\rho > \mathsf{D}_\rho(\infty)$ by Assumption (iii), a strictly feasibly point exists. Hence, we obtain by strong duality (see, e.g., [16, Theorem 8.6.1])

$$\begin{aligned}\mathsf{D}_{\rho,\tilde{\rho}}(\infty, D_\rho) &= \min_{\eta \geq 0} \eta D_\rho + \sup_{Q \in \mathcal{P}(\mathcal{X}\times\mathcal{Y}): Q_{\mathcal{X}} = P} \mathbb{E}_Q(\tilde{\rho} - \eta\rho) \\ &\leq \min_{\eta \geq 0} \eta D_\rho + \mathbb{E}_P g.\end{aligned}$$

As the last step, we show that we have equality in (16). To this end, we have to construct a $Q \in \mathcal{P}(\mathcal{X}\times\mathcal{Y})$ such that $Q_{\mathcal{X}} = P$ and $\mathbb{E}_Q f$ is arbitrarily close to $\mathbb{E}_P g$. Given any positive simple function $0 \leq s \leq g$, i.e., $s = \sum_{j=1}^J \beta_j \mathbb{1}_{B_j}$ for finite measurable partition $\{B_j\}_{j=1}^J$ of $\mathcal{X}$. Let $C_j \subset B_j$ be compact and such that $P(C_j) \geq P(B_j) - \varepsilon/J$ for all $j \in \{1, \ldots, J\}$. Since the $\{B_j\}_{j=1}^J$ are disjoint, we have $P(\cup_{j=1}^J C_j) \geq 1 - \varepsilon$. For each $x \in C_j$ and any $\delta > 0$, there exists a $y(x)$ such that

$$f(x, y(x)) \geq g(x) - \delta \geq s(x) - \delta.$$

By continuity of $f$ and since $s$ is constant on $B_j$, for each $x \in C_j \cap B_j$ there exists a open neighborhood $G_j(x)$ of $x$ such that

$$f(\tilde{x}, y(x)) \geq s(x) - 2\delta = s(\tilde{x}) - 2\delta$$

for every $\tilde{x} \in G_j(x)$. Since $C_j \subset \cup_{x \in B_j} G_j(x)$, and since $C_j$ is compact, there exists a finite subcover, say $\{G_j(x)\}_{x \in \widetilde{C}_j}$ for some finite set $\widetilde{C}_j \subset C_j$. Construct a finite measurable partition $\{E_k\}_{k=1}^K$ of $\cup_{j=1}^J C_j$ such that for each $k$ we have $E_k \subset G_j(x) \cap B_j$ for some $j$ and some $x \in \widetilde{C}_j$. Call $x_k$ the $x \in \widetilde{C}_j$ corresponding to $E_k$.

Define

$$Y \triangleq \begin{cases} y_0 & \text{if } X \in (\cup_{j=1}^J C_j)^c, \\ y(x_k) & \text{if } X \in E_k. \end{cases}$$

Since each $E_k$ is measurable, this is a random variable. Let $Q$ be the distribution of $(X, Y)$. We have

$$\begin{aligned}\mathbb{E}_Q f &= \sum_{k=1}^K \mathbb{E}_Q(f; E_k) + \mathbb{E}_Q(f(X,Y); (\cup_{j=1}^J C_j)^c) \\ &\geq \sum_{k=1}^K \mathbb{E}_P(f(X, y(x_k)); E_k) - \eta \mathbb{E}_P(\rho(X, y_0); (\cup_{j=1}^J C_j)^c) \\ &\geq \sum_{k=1}^K \mathbb{E}_P(s(X) - 2\delta; E_k) - \eta \mathbb{E}_P(\rho(X, y_0); (\cup_{j=1}^J C_j)^c) \\ &= \mathbb{E}_P s(X) - \mathbb{E}_P(\eta\rho(X, y_0) + s(X); (\cup_{j=1}^J C_j)^c) - 2\delta.\end{aligned}$$

Recall that $P(\cup_{j=1}^J C_j) \geq 1 - \varepsilon$. Since

$$0 \leq \mathbb{E}_P(\eta\rho(X, y_0) + s(X)) \leq \eta \mathbb{E}_P \rho(X, y_0) + \max_{j \in \{1,\ldots,J\}} \beta_j < \infty$$

by Assumption (ii), we can choose $\varepsilon$ small enough such that

$$E_P(\eta\rho(X,y_0)+s(X);(\cup_{j=1}^J C_j)^c)\le\delta.$$

With this

$$\mathbb{E}_Q f\ge\mathbb{E}_P s-3\delta. \tag{17}$$

Since $g$ is a measurable function, we can choose simple functions $s_i\le g$ such that $\lim_{i\to\infty}\mathbb{E}_P s_i=\mathbb{E}_P g$. In light of (17), this implies that

$$\sup_{Q\in\mathcal{P}(\mathcal{X}\times\mathcal{Y}):Q_\mathcal{X}=P}\mathbb{E}_Q f=\mathbb{E}_P g,$$

concluding the proof.

### *D. Proof of Theorem 4*

$\mathsf{D}_\rho(\cdot)$ is convex [15, Lemma 10.6.1] and hence continuous except for possibly at the boundary. By Assumption (iii), $\mathsf{D}_\rho(0)<\infty$, and by Assumption (i), $R>0$. Thus $\mathsf{D}_\rho(\cdot)$ is continuous at $R$. Therefore, since $D_\rho>\mathsf{D}_\rho(R)$ by Assumption (ii), and since $0<R<\infty$ by Assumption (i), there exists $\varepsilon>0$ such that $D_\rho-2\varepsilon\ge\mathsf{D}_\rho(R-\varepsilon)$. Hence by the definition of $\mathsf{D}_\rho(R)$, there exists $Q\in\mathcal{P}(\mathcal{X}\times\mathcal{Y})$ such that $Q_\mathcal{X}=P$, $I(Q)\le R-\varepsilon$ and $\mathbb{E}_Q\rho\le D_\rho-\varepsilon$.

Let $g_k:\mathcal{X}\to\mathcal{Y}$ be defined by

$$g_k(x)\triangleq\begin{cases}y_k^* & \text{if } x\in A_k,\\ y_0 & \text{else}.\end{cases}$$

Set $Y_k\triangleq g_k(X)$ and let $W_k$ be the distribution of $(X,Y_k)$. Set $\widetilde{Q}_k\triangleq(1-\alpha)Q+\alpha W_k$ for some $\alpha\in[0,1]$. Clearly both $W_k$ and $\widetilde{Q}_k$ have $\mathcal{X}$ marginal $P$. Mutual information is convex in the conditional distribution [15, Corollary 5.5.5], and thus

$$I(\widetilde{Q}_k)\le(1-\alpha)I(Q)+\alpha I(W_k)\le I(Q)+\alpha I(W_k).$$

We have $I(W_k)\le\log(2)<1$, and hence for $\alpha\le\varepsilon$

$$I(\widetilde{Q}_k)\le I(Q)+\varepsilon\le R. \tag{18}$$

Moreover, by Assumption (iii)

$$\begin{aligned}\mathbb{E}_{\widetilde{Q}_k}\rho&\le\mathbb{E}_Q\rho+\alpha\mathbb{E}_{W_k}\rho\\&\le D_\rho-\varepsilon+\alpha\big(\mathbb{E}_P(\rho(X,y_0);A_k^c)+\mathbb{E}_P(\rho(X,y_k^*);A_k)\big)\\&\le D_\rho-\varepsilon+\alpha\big(D_0+\mathbb{E}_P(\rho(X,y_k^*);A_k)\big).\end{aligned}$$

Setting

$$\alpha\triangleq\frac{\varepsilon}{1+D_0+\mathbb{E}_P(\rho(X,y_k^*);A_k)},$$

this becomes

$$\mathbb{E}_{\widetilde{Q}_k}\rho\le D_\rho. \tag{19}$$

Note that $\alpha\le\varepsilon$ as needed in (18), and $\alpha>0$ since $\mathbb{E}_P(\rho(X,y_k^*);A_k)<\infty$ by Assumption (iv).

Finally

$$\begin{aligned}\mathbb{E}_{\widetilde{Q}_k}\tilde{\rho}&\ge\alpha\mathbb{E}_{W_k}\tilde{\rho}\\&\ge\alpha\mathbb{E}_P(\tilde{\rho}(X,y_k^*);A_k)\\&=\varepsilon\frac{E_P(\tilde{\rho}(X,y_k^*);A_k)}{1+D_0+\mathbb{E}_P(\rho(X,y_k^*);A_k)},\end{aligned}$$

and

$$\mathbb{E}_P(\rho(X,y_k^*);A_k) = \mathbb{E}_P\Big(\frac{\rho(X,y_k^*)}{\tilde\rho(X,y_k^*)}\tilde\rho(X,y_k^*);A_k\Big)$$
$$\le E_P(\tilde\rho(X,y_k^*);A_k)\bigg(\sup_{x\in A_k}\frac{\rho(x,y_k^*)}{\tilde\rho(x,y_k^*)}\bigg).$$

Hence by Assumption (iv),

$$\mathbb{E}_{\tilde Q_k}\tilde\rho \ge \varepsilon\bigg(\frac{1+D_0}{\mathbb{E}_P(\tilde\rho(X,y_k^*);A_k)} + \sup_{x\in A_k}\frac{\rho(x,y_k^*)}{\tilde\rho(x,y_k^*)}\bigg)^{-1}$$
$$\ge \varepsilon\bigg(\frac{1+D_0}{P(A_k)\inf_{x\in A_k}\tilde\rho(x,y_k^*)} + \sup_{x\in A_k}\frac{\rho(x,y_k^*)}{\tilde\rho(x,y_k^*)}\bigg)^{-1} \to \infty \tag{20}$$

as $k\to\infty$.

Combining (18), (19), and (20), we get $\mathsf{D}_{\rho,\tilde\rho}(R,D_\rho)=\infty$.

### *E. Proof of Theorem 5*

$\mathsf{D}_{\rho,\tilde\rho}(\mathsf{R}_\rho(D_\rho),D_\rho)\ge 0$ if and only if the set of all $Q\in\mathcal{P}(\mathcal{X}\times\mathcal{Y})$ such that $Q_{\mathcal{X}}=P$, $I(Q)\le\mathsf{R}_\rho(D_\rho)$, $\mathbb{E}_Q\rho\le D_\rho$ is non empty. By definition $\mathsf{R}_\rho(D_\rho)=\inf I(Q)$, where the infimum is taken over all $Q\in\mathcal{P}(\mathcal{X}\times\mathcal{Y})$ such that $Q_{\mathcal{X}}=P$ and $\mathbb{E}_Q\rho\le D_\rho$. Hence $\mathsf{D}_{\rho,\tilde\rho}(\mathsf{R}_\rho(D_\rho),D_\rho)\ge 0$ if and only if this last infimum is attained (i.e., a minimizing $Q$ exists). By Theorem 2.2 (and the remark following its proof) in [17], this is the case when $\rho$ is continuous, $D_\rho>0$, and the set of all $Q$ over which the infimum is taken is tight[3].

From this, we only have to show tightness to prove the first part of the theorem. $\mathbb{E}_Q\rho\le D_\rho$ implies that

$$D_\rho\ge\mathbb{E}_Q\rho\ge Q(K_k\times M_k^c)\inf_{x\in K_k,y\in M_k^c}\rho(x,y),$$

and thus

$$Q(K_k\times M_k) = P(K_k) - Q(K_k\times M_k^c)$$
$$\ge P(K_k) - D_\rho/\inf_{x\in K_k,y\in M_k^c}\rho(x,y)\to 1$$

as $k\to\infty$. Since the sets $K_k\times M_k$ are compact, this shows tightness and proves the first part of the theorem.

The proof of the second part adapts an argument from [17, Theorem 2.2]. Note that since $\mathsf{D}_{\rho,\tilde\rho}(\mathsf{R}_\rho(D_\rho)+r,D_\rho)<\infty$ for some $r>0$ (and hence, by concavity, for all $r>0$), for every $\varepsilon>0$ and all $i\ge 1$ there exists $Q_i\in\mathcal{P}(\mathcal{X}\times\mathcal{Y})$ with $\mathcal{X}$ marginal $P$ such that

$$I(Q_i)\le\mathsf{R}_\rho(D_\rho)+1/i,$$
$$\mathbb{E}_{Q_i}\rho\le D_\rho,$$
$$\mathbb{E}_{Q_i}\tilde\rho\ge\mathsf{D}_{\rho,\tilde\rho}(\mathsf{R}_\rho(D_\rho)+1/i,D_\rho)-\varepsilon.$$

Since the set of all feasible distributions is tight as shown above, this implies that $\{Q_i\}_{i\ge1}$ contains a weakly convergent subsequence[4], and we may assume without loss of generality that $Q_i\Rightarrow Q$ for some $Q\in\mathcal{P}(\mathcal{X}\times\mathcal{Y})$. Using exactly the same argument as in [17, Theorem 2.2], we have

$$\mathsf{R}_\rho(D_\rho)\ge\liminf_{i\to\infty}I(Q_i)\ge I(Q) \tag{21}$$

[3]The set of distributions $\mathcal{Q}\subset\mathcal{P}(\mathcal{X}\times\mathcal{Y})$ is tight if there exists compact sets $A_k\subset\mathcal{X}\times\mathcal{Y}$ such that $\sup_{Q\in\mathcal{Q}}Q(A_k^c)\to 0$ as $k\to\infty$.

[4]$Q_i$ converges weakly to $Q$ (denoted by $Q_i\Rightarrow Q$) if $\lim_{i\to\infty}\mathbb{E}_{Q_i}g=\mathbb{E}_Q g$ for all bounded and continuous functions $g\in\mathcal{X}\times\mathcal{Y}\to\mathbb{R}$. An equivalent definition for $Q_i\Rightarrow Q$ is that $\liminf_{i\to\infty}Q_i(A)\ge Q(A)$ for all open sets $A\subset\mathcal{X}\times\mathcal{Y}$ (see [18, Theorem 2.1]). If $Z_i\sim Q_i$ and $Z\sim Q$, we write $Z_i\Rightarrow Z$ if $Q_i\Rightarrow Q$.

and

$$D_\rho \geq \liminf_{i\to\infty} \mathbb{E}_{Q_i}\rho \geq \mathbb{E}_Q\rho. \tag{22}$$

Finally, since $\tilde{\rho}$ is continuous, we have $\tilde{\rho}(X,Y_i) \Rightarrow \tilde{\rho}(X,Y)$, where $(X,Y_i)\sim Q_i$ and $(X,Y)\sim Q$. As

$$\sup_{i\geq 1} \mathbb{E}\tilde{\rho}(X,Y_i)^a \leq c + \sup_{i\geq 1}\mathbb{E}\rho(X,Y_i) \leq c + D_\rho < \infty,$$

$\{\tilde{\rho}(X,Y_i)\}_{i\geq 1}$ is uniformly integrable. Therefore by [18, Theorem 3.5]

$$\lim_{i\to\infty} \mathsf{D}_{\rho,\tilde{\rho}}(\mathsf{R}_\rho(D_\rho)+1/i, D_\rho) - \varepsilon \leq \lim_{i\to\infty}\mathbb{E}\tilde{\rho}(X,Y_i) = \mathbb{E}\tilde{\rho}(X,Y) = \mathbb{E}_Q\tilde{\rho}. \tag{23}$$

Since $\varepsilon > 0$ is arbitrary, (21), (22), and (23), imply that

$$\mathsf{D}_{\rho,\tilde{\rho}}(\mathsf{R}_\rho(D_\rho), D_\rho) \geq \mathsf{D}_{\rho,\tilde{\rho}}(\mathsf{R}_\rho(D_\rho)+, D_\rho).$$

As $\mathsf{D}_{\rho,\tilde{\rho}}(\cdot, D_\rho)$ is increasing, we also have

$$\mathsf{D}_{\rho,\tilde{\rho}}(\mathsf{R}_\rho(D_\rho), D_\rho) \leq \mathsf{D}_{\rho,\tilde{\rho}}(\mathsf{R}_\rho(D_\rho)+, D_\rho),$$

concluding the proof of the second part of the theorem.

### *F. Proof of Theorem 6*

Since $\mathcal{Q}_2(D_\rho, D_{\tilde{\rho}}) \subset \mathcal{Q}_1(D_\rho, D_{\tilde{\rho}})$, it is enough to show that for every $\delta > 0$

$$\inf_{Q\in\mathcal{Q}_1(D_\rho,D_{\tilde{\rho}})} I(Q) \geq \inf_{Q\in\mathcal{Q}_2(D_\rho+\delta,D_{\tilde{\rho}}-\delta)} I(Q).$$

For some $\nu > 0$, choose $Q\in\mathcal{Q}_1(D_\rho, D_{\tilde{\rho}})$ such that

$$I(Q) \leq \inf_{Q\in\mathcal{Q}_1(D_\rho,D_{\tilde{\rho}})} I(Q) + \nu.$$

Fix $\varepsilon > 0$ and let $Z$ be uniformly distributed on $(-\varepsilon,\varepsilon)^m$ and independent of $X, Y$. Define $\widetilde{Y} \triangleq Y + Z$ and let $Q_\varepsilon$ be the distribution of $(X,\widetilde{Y})$ when $(X,Y)\sim Q$. Note that by Assumption (iii), $Q_\varepsilon \ll \lambda_{\mathbb{R}^m\times\mathbb{R}^m}$ whenever $\varepsilon > 0$ and that $Q_0 = Q$. By the data processing inequality

$$I(Q_\varepsilon) \leq I(Q) \leq \inf_{Q\in\mathcal{Q}_1(D_\rho,D_{\tilde{\rho}})} I(Q) + \nu. \tag{24}$$

We now show that $Q_\varepsilon \Rightarrow Q$ as $\varepsilon\to 0$ (i.e., that $Q_\varepsilon$ converges weakly[4] to $Q$). For this, it suffices to show that for every open $G\subset\mathcal{X}\times\mathcal{Y}$ we have $\liminf_{\varepsilon\to 0} Q_\varepsilon(G) \geq Q(G)$. Define

$$G_\varepsilon \triangleq \{(x,y)\in\mathcal{X}\times\mathcal{Y} : (x, y+z)\in G \ \forall z\in\mathbb{R}^m \text{ with } \|z\|_\infty < \varepsilon\}.$$

Since $(X,Y)\in G_\varepsilon$ implies $(X,\widetilde{Y})\in G$, we have $Q_\varepsilon(G)\geq Q(G_\varepsilon)$. Since $G$ is open, we have $\mathbb{1}_{G_\varepsilon}\to\mathbb{1}_G$ pointwise as $\varepsilon\to 0$, and hence by Fatou's lemma

$$\liminf_{\varepsilon\to 0} Q_\varepsilon(G) \geq \liminf_{\varepsilon\to 0} Q(G_\varepsilon) \geq Q(G).$$

Thus $Q_\varepsilon \Rightarrow Q$ as $\varepsilon\to 0$.

By continuity of $\tilde{\rho}$ (Assumption (i)), we get by weak convergence for every $b\geq 0$

$$\mathbb{E}_{Q_\varepsilon}\tilde{\rho} \geq \mathbb{E}_{Q_\varepsilon}\min\{\tilde{\rho}, b\} \to \mathbb{E}_Q\min\{\tilde{\rho}, b\}, \tag{25}$$

as $\varepsilon\to 0$. Assuming $\mathbb{E}_Q\tilde{\rho} < \infty$, choose $b$ such that $E_Q\min\{\tilde{\rho},b\} \geq \mathbb{E}_Q\tilde{\rho} - \delta/2$. Then there exists $\varepsilon_1 > 0$ such that for $\varepsilon\leq\varepsilon_1$, we have by (25)

$$\begin{aligned}\mathbb{E}_{Q_\varepsilon}\tilde{\rho} &\geq \mathbb{E}_Q\min\{\tilde{\rho},b\} - \delta/2\\ &\geq \mathbb{E}_Q\tilde{\rho} - \delta\\ &\geq D_{\tilde{\rho}} - \delta.\end{aligned} \tag{26}$$

Since $D_{\tilde{\rho}} < \infty$, this last conclusion follows by a similar argument if $\mathbb{E}_Q\tilde{\rho} = \infty$.

Moreover, by Assumption (ii)

$$\begin{aligned}\mathbb{E}_{Q_\varepsilon}\rho &= \mathbb{E}_{Q_\varepsilon}(\rho; A^c) + \mathbb{E}_{Q_\varepsilon}(\rho; A)\\ &\leq \mathbb{E}_{Q_\varepsilon}(\rho; A^c) + \mathbb{E}_Q\big(\sup_{z:\|z\|_\infty\leq\varepsilon}\rho(X, Y+z); A\big)\\ &\leq \mathbb{E}_{Q_\varepsilon}(\rho; A^c) + c\mathbb{E}_Q(\rho; A).\end{aligned} \tag{27}$$

Now note that Assumption (ii) holds also as we increase $a$. Since $\mathbb{E}_Q\rho \leq D_\rho$, we have $E_Q(\rho; A) \to 0$ as $a \to \infty$. Hence there exists $a$ such that Assumption (ii) holds and $c\mathbb{E}_Q(\rho; A) \leq \delta/2$. For this $a$, we can continue (27) as

$$\begin{aligned}\mathbb{E}_{Q_\varepsilon}\rho &\leq \mathbb{E}_{Q_\varepsilon}(\rho; A^c) + \delta/2\\ &\leq \mathbb{E}_{Q_\varepsilon}\min\{\rho, a\} + \delta/2.\end{aligned}$$

By continuity of $\rho$ (Assumption (i)) and weak convergence of $Q_\varepsilon$, $\mathbb{E}_{Q_\varepsilon}\min\{\rho, a\} \to \mathbb{E}_Q\min\{\rho, a\}$ as $\varepsilon \to 0$. Hence there exists $0 < \varepsilon_2 \leq \varepsilon_1$ such that for $0 < \varepsilon \leq \varepsilon_2$, we have

$$\begin{aligned}\mathbb{E}_{Q_\varepsilon}\rho &\leq \mathbb{E}_Q\min\{\rho, a\} + \delta\\ &\leq \mathbb{E}_Q\rho + \delta\\ &\leq D_\rho + \delta.\end{aligned} \tag{28}$$

Combining (24), (26), and (28), we obtain for $0 < \varepsilon \leq \varepsilon_2$ that $Q_\varepsilon \in \mathcal{Q}_2(D_\rho + \delta, D_{\tilde{\rho}} - \delta)$ and

$$I(Q_\varepsilon) \leq \inf_{Q\in\mathcal{Q}_1(D_\rho, D_{\tilde{\rho}})} I(Q) + \nu.$$

Since $\nu > 0$ is arbitrary, this shows that

$$\inf_{Q\in\mathcal{Q}_2(D_\rho+\delta, D_{\tilde{\rho}}-\delta)} I(Q) \leq \inf_{Q\in\mathcal{Q}_1(D_\rho, D_{\tilde{\rho}})} I(Q),$$

proving the first part of the theorem.

The second part follows directly from continuity of

$$\inf_{Q\in\mathcal{Q}_2(D_\rho, D_{\tilde{\rho}})} I(Q).$$

### *G. Proof of Theorem 7*

Let $Q \in \mathcal{Q}_2(D_\rho, D_{\tilde{\rho}})$ and $(X, Y) \sim Q$. Then $h(X)$ and $h(X|Y)$ are well defined and since $h(X)$ is finite, we have $I(Q) = h(X) - h(X|Y)$. Therefore

$$\begin{aligned}I(Q) &= h(X) - h(X|Y)\\ &= h(X) - h(X-Y|Y)\\ &\geq h(X) - h(X-Y)\\ &\geq h(X) - \sup_{Z:\mathbb{E}\rho(Z)\leq D_\rho, \mathbb{E}\tilde{\rho}(Z)\geq D_{\tilde{\rho}}} h(Z),\end{aligned}$$

with equality if there exists $Z = X - Y$ independent of $Y$ and achieving the supremum. By [19, Theorem 3.2], if there exist $\alpha \in \mathbb{R}, \eta, \tilde{\eta} \in \mathbb{R}_+$ such that $f: \mathbb{R}^m \to \mathbb{R}_+$ defined by

$$f(z) \triangleq \exp\big(-\alpha - \eta\rho(z) + \tilde{\eta}\tilde{\rho}(z)\big)$$

satisfies

$$\int f(z)dz = 1,$$
$$\int \rho(z)f(z)dz = D_\rho,$$
$$\int \tilde{\rho}(z)f(z)dz = D_{\tilde{\rho}},$$

then $f$ is the density of the maximizing $Z$, and

$$h(Z) = \alpha + \eta D_\rho - \tilde{\eta} D_{\tilde{\rho}}.$$

Thus in this case

$$\inf_{Q \in \mathcal{Q}_2(D_\rho, D_{\tilde{\rho}})} I(Q) \geq \max\big\{0, h(X) - \alpha - \eta D_\rho + \tilde{\eta} D_{\tilde{\rho}}\big\}.$$

*H. Proof of Corollary 8*

Let $\varepsilon, \delta > 0$ be small enough such that

$$D_{\widetilde{\Gamma}} < \mathsf{D}_{\Gamma,\widetilde{\Gamma}}\big(R, \{\Delta_\rho - \delta\}\big) - 2\varepsilon$$

and

$$\inf_{\rho \in \Gamma} \Delta_\rho > \delta. \tag{29}$$

For every $\rho \in \Gamma$, we have

$$\begin{aligned} D_{\widetilde{\Gamma}} &< \mathsf{D}_{\Gamma,\widetilde{\Gamma}}\big(R, \{\Delta_\rho - \delta\}\big) - 2\varepsilon \\ &\leq \sup_{\tilde{\rho} \in \widetilde{\Gamma}} \mathsf{D}_{\rho,\tilde{\rho}}(R, \mathsf{D}_\rho(R) + \Delta_\rho - \delta) - 2\varepsilon \\ &\leq \mathsf{D}_{\rho,\tilde{\rho}}(R, \mathsf{D}_\rho(R) + \Delta_\rho - \delta) - \varepsilon, \end{aligned}$$

for some $\tilde{\rho} \in \widetilde{\Gamma}$. By (29), $\Delta_\rho - \delta > 0$, and hence $\mathsf{D}_{\rho,\tilde{\rho}}(\cdot, \mathsf{D}_\rho(R) + \Delta_\rho - \delta)$ is continuous at $R$. Therefore, by choosing $\delta$ small enough, we have

$$\mathsf{D}_{\rho,\tilde{\rho}}(R, \mathsf{D}_\rho(R) + \Delta_\rho - \delta) - \varepsilon \leq \mathsf{D}_{\rho,\tilde{\rho}}(R - \delta, \mathsf{D}_\rho(R) + \Delta_\rho - \delta).$$

Hence for this $\tilde{\rho}$, Theorem 1 guarantees the existence of a sequence of source codes $\{f_n\}_{n\geq 1}$ such that

$$\begin{aligned} \lim_{n\to\infty} \frac{1}{n} \log |f_n(\mathcal{X}^n)| &\leq R, \\ \limsup_{n\to\infty} \mathbb{E}\rho_n(X^n, f_n(X^n)) &\leq \mathsf{D}_\rho(R) + \Delta_\rho, \\ \liminf_{n\to\infty} \mathbb{E}\tilde{\rho}_n(X^n, f_n(X^n)) &\geq D_{\widetilde{\Gamma}}. \end{aligned}$$

This proves part a of the theorem.

Part b follows similarly.

### I. Proof of Corollary 9

Choose $\rho \in \Gamma$ such that

$$\sup_{\tilde{\rho} \in \widetilde{\Gamma}} \mathsf{D}_{\rho,\tilde{\rho}}(R, \mathsf{D}_\rho(R) + \Delta_\rho) \leq \mathsf{D}_{\Gamma,\widetilde{\Gamma}}\big(R, \{\Delta_\rho\}\big) + \delta.$$

For any $\tilde{\rho} \in \widetilde{\Gamma}$, we have by Theorem 2

$$E\tilde{\rho}_n(X^n, f_n(X^n)) \leq \mathsf{D}_{\rho,\tilde{\rho}}(R+, \mathsf{D}_\rho(R) + \Delta_\rho). \tag{30}$$

Since $\Delta_\rho > 0$, $\mathsf{D}_{\rho,\tilde{\rho}}(\cdot, \mathsf{D}_\rho(R) + \Delta_\rho)$ is continuous at $R$. Hence

$$\begin{aligned}
\mathsf{D}_{\rho,\tilde{\rho}}(R+, \mathsf{D}_\rho(R) + \Delta_\rho) &= \mathsf{D}_{\rho,\tilde{\rho}}(R, \mathsf{D}_\rho(R) + \Delta_\rho) \\
&\leq \sup_{\tilde{\rho} \in \widetilde{\Gamma}} \mathsf{D}_{\rho,\tilde{\rho}}(R, \mathsf{D}_\rho(R) + \Delta_\rho) \\
&\leq \mathsf{D}_{\Gamma,\widetilde{\Gamma}}\big(R, \{\Delta_\rho\}\big) + \delta.
\end{aligned}$$

This proves part a of the theorem.

Part b follows similarly.

### J. Proof of Proposition 10

$$\begin{aligned}
\mathsf{D}_{\rho_1,\rho_3}(R, D) &= \sup_{\substack{Q \in \mathcal{P}(\mathcal{X}\times\mathcal{Y}): Q_{\mathcal{X}} = P, \\ I(Q) \leq R, \mathbb{E}_Q \rho_1 \leq D}} \mathbb{E}_Q \rho_3 \\
&\leq \mathsf{D}_{\rho_1,\rho_2}(R, D) + \sup_{Q \in \mathcal{P}(\mathcal{X}\times\mathcal{Y}): Q_{\mathcal{X}} = P} \mathbb{E}_Q(\rho_3 - \rho_2) \\
&\leq \mathsf{D}_{\rho_1,\rho_2}(R, D) + \mathbb{E}_P(\sup_{y \in \mathcal{Y}} \rho_3(X, y) - \rho_2(X, y)).
\end{aligned}$$

And

$$\begin{aligned}
\mathsf{D}_{\rho_1,\rho_3}(R, D) &= \sup_{\substack{Q \in \mathcal{P}(\mathcal{X}\times\mathcal{Y}): Q_{\mathcal{X}} = P, \\ I(Q) \leq R, \mathbb{E}_Q \rho_1 \leq D}} \mathbb{E}_Q \rho_3 \\
&\geq \mathsf{D}_{\rho_1,\rho_2}(R, D) - \sup_{Q \in \mathcal{P}(\mathcal{X}\times\mathcal{Y}): Q_{\mathcal{X}} = P} \mathbb{E}_Q|\rho_3 - \rho_2| \\
&\geq \mathsf{D}_{\rho_1,\rho_2}(R, D) - \mathbb{E}_P \sup_{y \in \mathcal{Y}} |\rho_3(X, y) - \rho_2(X, y)|.
\end{aligned}$$

## IV. Conclusion

In this paper, we investigated the problem of source coding with mismatched distortion measures. We derived a single-letter characterization $\mathsf{D}_{\rho,\tilde{\rho}}(R, D_\rho)$ of the best distortion level with respect to $\tilde{\rho}$ that can be guaranteed for any source code of rate $R$ designed for distortion level $D_\rho$ with respect to $\rho$. We also derived a single-letter characterization $\mathsf{D}_{\rho,\tilde{\rho}}(\infty, D_\rho)$ of the best distortion guarantee independent of the rate $R$ of the source code. We then looked at properties of $\mathsf{D}_{\rho,\tilde{\rho}}(R, D_\rho)$, characterizing its behavior for $R > \mathsf{R}_\rho(D_\rho)$ and on the boundary. We finally considered the problem of choosing a representative $\rho \in \Gamma$ of $\tilde{\rho}$.

## V. Acknowledgment

The authors would like to thank Ram Zamir for helpful discussions.